\documentclass[iop]{emulateapj}
\usepackage{apjfonts}

\usepackage{graphicx,longtable}
\usepackage{array}
\usepackage{amsmath}
\usepackage{amssymb}
\usepackage{subfigure}

\submitted{}

\begin{document}
\title{Quantum Statistical Corrections to Astrophysical Photodisintegration Rates}
\author{G.~J. Mathews\altaffilmark{1,3}, Yamac Pehlivan\altaffilmark{2,3}, Toshitaka
Kajino\altaffilmark{3,4}, A.~B. Balantekin\altaffilmark{5},
and Motohiko Kusakabe\altaffilmark{6}}

\altaffiltext{1}{Center for Astrophysics, Department of Physics, University of Notre Dame, Notre Dame, IN 46556, USA\\
{\tt gmathews@nd.edu}}
\altaffiltext{2}{Mimar Sinan Fine Arts University Besiktas, Istanbul 34349, Turkey\\
{\tt yamac@physics.wisc.edu}}
\altaffiltext{3}{National Astronomical Observatory of Japan 2-21-1
Osawa, Mitaka, Tokyo, 181-8588, Japan\\
{\tt kajino@nao.ac.jp}}
\altaffiltext{4}{Department of Astronomy, School of Science, the
University of Tokyo, 7-3-1 Hongo, Bunkyo-ku, Tokyo, 113-0033, Japan
}
\altaffiltext{5}{Department of Physics, University of Wisconsin, Madison, WI 53706, USA\\
{\tt baha@physics.wisc.edu}}
\altaffiltext{6}{Institute for Cosmic Ray Research, University of Tokyo,
Kashiwa, Chiba 277-8582, Japan\\
{\tt kusakabe@icrr.u-tokyo.ac.jp}}
\begin{abstract}
Tabulated rates for astrophysical photodisintegration reactions make use of Boltzmann statistics for the photons involved as well as the interacting nuclei.  Here we derive analytic  corrections for the Planck-spectrum quantum statistics of the photon energy distribution.  These corrections can be deduced directly from the detailed-balance condition without the assumption of equilibrium as long as the photons are represented by a Planck spectrum.  Moreover we show that these corrections affect not only the photodisintegration rates but also modify the conditions of nuclear statistical equilibrium as represented in the Saha equation. We deduce  new  analytic corrections to the classical
Maxwell-Boltzmann statistics  which can easily be added to the reverse reaction rates of existing reaction network tabulations.
We show that the effects of quantum statistics, though generally quite small,  always tend to speed up photodisintegration rates and  are largest for nuclei and environments for which  $Q/kT \sim 1$. As an illustration, we examine  possible effects of these corrections on  the $r$-process, the $rp$-process, explosive silicon burning, the $\gamma$-process and big bang nucleosynthesis. We find that in most cases one is quite justified in neglecting these corrections.  The correction is largest for  reactions near the drip line for an $r$-process with very high neutron density, or an $rp$-process at high-temperature. 
\end{abstract}
\medskip
\maketitle

\vskip 1.3cm



\section{Introduction}

The capture of a projectile particle by a nucleus followed
by the emission of a photon is called radiative capture. 
The inverse process is called photodisintegration. Both types
of reactions play important roles in  stellar and big bang
nucleosynthesis \citep*{rol90,smi93,wag67,fow67,cla68,Iliadis07}. 
Photodisintegration in astrophysical environments 
 often involves a thermal distribution of photons that  excite nuclei above the particle emission threshold.  
Above a temperature of $T \sim 10^9$ K, photodissociation can become a dominant process in the reaction flow and the photo-ejected nucleons 
can be captured by other nuclei leading to photodisintegration rearrangement  as can happen for example during core or explosive  oxygen or  silicon burning \citep{cla68, Iliadis07}.

In determining the rate for photodisintegration reactions, however, one should take into account
various factors arising from dealing with  photons in a two body problem.
One of them is the fact that photons are massless  bosons, and hence, obey Planckian statistics.
Historically, photodisintegration reactions have been  treated with Maxwell-Boltzmann statistics, both because it is usually an excellent approximation (cf. \cite{Rauscher95,Iliadis07}) and because this assumption simplifies the determination of the photodisintegration rates.  However, low-energy photons are more correctly represented by a  Planck distribution.
As discussed below, if the ratio of the capture $Q$-value to the temperature $Q/kT$
is large then the use of a Maxwell-Boltzmann distribution for the photons is a good approximation. On the other hand, for $Q/kT \sim 1$ (e.g., if one is interested in the photo-ejection of loosely bound particles from nuclei), then the
effects of quantum statistics become more relevant.
We show here that  modified thermal photodisintegration rates can be written in the form of a small analytic correction to
the tabulated reaction rates obtained with Maxwell-Boltzmann statistics.  Furthermore, these correction factors are nearly independent of the nuclear cross sections and to leading order only depend upon the reaction $Q$-value,  the Gamow energy (for the charged-particle nonresonant part), and the resonance energy (for the resonant part).

\subsection{Photonuclear Reactions}

For a reaction (not necessarily in equilibrium) of the form 
\begin{equation}\label{Reaction}
1+2 \rightarrow 3+\gamma~~, ~~3+\gamma \rightarrow  1 + 2~~,
\end{equation}
the forward and reverse reaction rates are given by
\begin{eqnarray}
r_{12}&=&n_{1}n_{2}\langle\sigma v\rangle_{12} = n_{1} \lambda_{12} \\
 r_{\gamma3}& = &n_{3}n_{\gamma}\langle\sigma c\rangle_{\gamma3}= n_{3} \lambda_{\gamma 3}~~.
\label{eq:Rates}
\end{eqnarray}
Here, $\langle\sigma v\rangle_{12}$ denotes the thermally
averaged reaction rate per particle pair for the capture reaction and is given by an integration over an appropriate velocity distribution $\phi(v)$,
\begin{equation}
\langle\sigma v\rangle_{12}=\int\sigma_{12}\left(v\right)\: v\:\phi(v)\: d^{3}v~~,
\end{equation}
where $v$ denotes the relative velocity of the nuclei $1$ and $2$.
For most  cases of interest in astrophysics,
the massive interacting particles  are non-degenerate (i.e., dilute) and non-relativistic
(i.e., their rest mass energy is large compared to $kT$). Hence, one can use a Maxwell-Boltzmann velocity distribution with
Newtonian kinetic energy because
\begin{equation}
\frac{1}{e^{\sqrt{p^{2}c^{2}+m^{2}c^{4}}/kT}\pm1}\simeq e^{-\sqrt{p^{2}c^{2}+m^{2}c^{4}}/kT}\propto e^{-\frac{mv^{2}}{2kT}}~~,
\end{equation}
where the exponential factor involving the rest mass energy drops out with
a proper normalization of the distribution function. Also note that
when both particles $1$ and $2$ obey Maxwell-Boltzmann
statistics, so does their relative velocity \citep{cla68}. Hence, the thermally
averaged capture rate per particle pair is given by
\begin{equation}
\langle\sigma v\rangle_{12}=\sqrt{\frac{8}{\pi\mu}}\left(kT\right)^{-\frac{3}{2}}\int_{0}^{\infty}\sigma_{12}\left(E\right)\: e^{-E/kT}\: E\: dE~~,\label{eq:Forward Reaction Rate}
\end{equation}
where  $\mu$ denotes the reduced mass for the particles
$1$ and $2$.

\begin{figure}
\begin{center}
\includegraphics[scale=0.2]{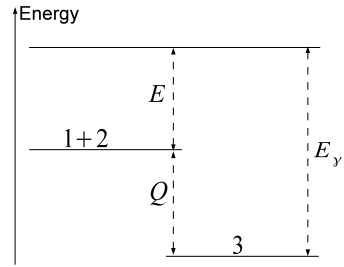}
\end{center}
\caption{Schematic illustration of the energy threshold for photodisintegration.}\label{fig:Energy Figure}
\end{figure}

In  the photodisintegration reaction rate, however, the relative velocity of the photon
with respect to the target nucleus is always the speed of light $c$. This
eliminates any dependence of the reaction rate on the velocity distribution
of the target nuclei \citep{thi98}. Therefore, the photonuclear reaction rate $\langle\sigma c\rangle_{\gamma3}$
becomes an integral of  the reaction cross section over a Planck energy
distribution for the  photons:
\begin{equation}
\langle\sigma c\rangle_{\gamma3}=\frac{1}{2 \zeta(3) \left(kT\right)^{3}}
\int_{\mbox{\footnotesize max}(0,Q)}^{\infty}\sigma_{\gamma3}\left(E_{\gamma}\right)\: c\:\frac{1}{e^{E_{\gamma}/kT}-1}\: E_{\gamma}^{2}\: dE_{\gamma}~~.
\label{eq:Inverse Reaction Rate}
\end{equation}
Using the fact that for a Planck distribution
\begin{equation}
n_\gamma = {16 \pi} { \zeta(3)}\biggl(\frac {kT}{h c}\biggr)^{3}~~,
\end{equation}
we can equivalently write Eq.~(\ref{eq:Inverse Reaction Rate}) in more familiar form
\begin{equation}
\lambda_{\gamma 3}=\frac{8 \pi}{(h c)^{3}}
\int_{\mbox{\footnotesize max}(0,Q)}^{\infty}\sigma_{\gamma3}\left(E_{\gamma}\right)\: c\:\frac{1}{e^{E_{\gamma}/kT}-1}\: E_{\gamma}^{2}\: dE_{\gamma}~~,
\label{eq:Inverse lambda}
\end{equation}
where $\zeta(3) = 1.20206$ is the Riemann zeta function and $E_{\gamma}$ denotes the photon energy.  The integration threshold is the $Q$-value of
the capture reaction (see Figure \ref{fig:Energy Figure}) or zero in the case of negative $Q$. 

\placefigure{fig:Energy Figure}

\subsection{Detailed Balance Condition}

It is  difficult to determine the cross section $\sigma_{\gamma3}$ directly from 
experiment. However,  the interaction between photons and matter is very
weak $(e^{2}/\hbar c\ll1)$ so that the reaction can be treated with first order
perturbation theory. In this case, the transition probabilities become
proportional to the matrix elements of the perturbing Hamiltonian and
the hermiticity of the perturbing Hamiltonian gives rise to a simple relation
between the capture and disintegration cross sections.  This is known
as the \emph{detailed balance equation} \citep{bla91}. For a reaction involving  a ground-state to ground-state transition for two nuclei with energy $E$, leading to a gamma ray with energy $E_{\gamma}=E+Q$ this is  given by
\begin{equation}
\sigma_{\gamma3}(E_{\gamma})=\frac{g_{1}g_{2}}{ g_{3}(1 + \delta_{12})}\frac{\mu c^2 E}{E_{\gamma}^{2}}\sigma_{12}(E)~~,
\label{eq:Detailed Balance}
\end{equation}
where  $g_{i} = 2 j_i + 1 $ are the spin degeneracy factors for the ground state of the nuclei and the Kronecker delta function accounts for the special case of indistinguishable interacting nuclei.
 Using this detailed balance equation, the photodisintegration  rate for a single-state transition can be related to the forward capture rate.

 Substituting Eq.~(\ref{eq:Detailed Balance})
into Eq.~(\ref{eq:Inverse Reaction Rate}) and also changing the variable from $E_{\gamma}$ to $E=E_{\gamma}-Q$ in the integration,
one can average over the velocity distribution of the ground-state interacting nuclei $\langle\sigma c\rangle_{\gamma3}$ as follows:
\begin{equation}
\langle\sigma c\rangle_{\gamma3}=\frac{\mu c^{3}}{2 \zeta(3) (kT)^{3}}\frac{g_{1}g_{2}}{g_{3} (1 + \delta_{12})}\int_{0}^{\infty}\sigma_{12}\left(E\right)\frac{1}{e^{\left(E+Q\right)/kT}-1}EdE.
\label{eq:Inverse Reaction Rate 2}
\end{equation}
At this point, one usually introduces the approximation: 
\begin{equation}
e^{\left(E+Q\right)/kT}-1 \approx e^{\left(E+Q\right)/kT}~~.
\label{approx}
\end{equation}
Here, we point out that by inserting this approximation and then correcting for it,  Eq.~(\ref{eq:Inverse Reaction Rate 2}) can  be rewritten in  the following exact form:
\begin{equation}
\langle\sigma c\rangle_{\gamma3} =  \bigl(1+R\bigr) {\langle\sigma v\rangle_{12}}  \biggl( \frac{ \sqrt{2 \pi}}{8 \zeta(3)} \biggr) 
\frac{g_{1}g_{2}}{g_{3} (1 + \delta_{12})} \left(\frac{\mu c^{2}}{kT}\right)^{3/2}e^{-Q/kT}~,
\label{eq:Ratio of Reaction Rates 3}
\end{equation}
where  $R$ is a small and dimensionless number which is formally given by
\begin{equation}
1 + R=\biggl[\frac{\int_{0}^{\infty}\sigma_{12}\left(E\right)(e^{\left(E+Q\right)/kT}-1)^{-1}EdE}{\int_{0}^{\infty}\sigma_{12}\left(E\right)e^{-\left(E+Q\right)/kT}EdE}\biggr]~~.
\label{eq:R-Formal}
\end{equation}

\subsection{Thermal Population of Excited States}

The generalization of Eq. (\ref{eq:Ratio of Reaction Rates 3}) to the average over  thermally populated states  among the initial and final nuclei is straightforward \citep{cla68, Iliadis07}.  One must first replace the ground state (g.s.) to g.s.~forward reaction cross section $\sigma_{1 2}$  with a weighted average over the thermal population  of states $\mu$ in the target nucleus 1, and also sum over all final states in product nucleus $3$. (Note that we only consider light particle $(p,\gamma)$, $(n,\gamma)$, or $(\alpha,\gamma)$, reactions for which we can ignore their excitation.)  Thus, the effective stellar thermal forward rate becomes
\begin{equation}
\langle \sigma v \rangle_{1 2}^* =  \frac{\sum_\mu g_{1 \mu} e^{-E_{1 \mu}/kT} 
\sum_\nu  \langle \sigma v \rangle_{1 2}^{\mu \rightarrow \nu}}{\sum_\mu g_{1 \mu} e^{-E_{1 \mu}/kT} }~~,
\end{equation}
which can also be written as
\begin{equation}
\langle \sigma v \rangle_{1 2}^* =  R_{t t} \langle \sigma v \rangle_{1 2} ~~,
\end{equation}
where the stellar enhancement factor $R_{t t}$ is defined by
\begin{equation}
R_{t t} = \frac{\sum_\mu g_{1 \mu} e^{-E_{1 \mu}/kT} 
\frac{ \sum_\nu \langle \sigma_{1 2}  v \rangle_{1 2}^{\mu \rightarrow \nu}}{\sum_\nu \langle \sigma_{1 2}  v \rangle_{1 2}^{g.s. \rightarrow \nu}}}{\sum_\mu g_{1 \mu} e^{-E_{1 \mu}/kT} }~~.
 \end{equation} 
Usually, tabulated thermonuclear reaction rates are given as the ground state rate and the stellar enhancement factor must be determined from a statistical model calculation as in \cite{Holmes76}, \cite{Woosley78}, \cite{Rauscher00} and \cite{Rauscher04}.

The thermally averaged photonuclear rate for a distribution of excited states in initial and final heavy nuclei then becomes
\begin{equation}
\langle \sigma c \rangle_{3 \gamma}^* = \frac{\sum_\nu g_{3 \nu} e^{-E_{3 \nu}/kT}
 \sum_\mu  \langle \sigma c \rangle_{3 \gamma}^{\nu \rightarrow \mu}}{\sum_\nu g_{3 \nu} e^{-E_{3 \nu}/kT} } ~~.
\label{sigmac-star}
\end{equation}

The generalization of the detailed balance condition of Eq.~(\ref{eq:Ratio of Reaction Rates 3}) is
\begin{eqnarray}
\langle \sigma c \rangle_{\gamma 3}^{\mu \rightarrow \nu} & = & \bigl(1+R_{\mu \nu} \bigr) {\langle\sigma v \rangle_{12}^{\nu \rightarrow \mu}}  \biggl( \frac{ \sqrt{2 \pi}}{8 \zeta(3)} \biggr) 
\frac{g_{1 \mu}g_{2}}{g_{3 \nu } (1 + \delta_{12})} \nonumber \\
&& \times  \left(\frac{\mu c^{2}}{kT}\right)^{3/2}e^{-Q_{\mu \nu}/kT}~~,
\label{eq:Ratio of Reaction Rates munu}
\end{eqnarray}
where $R_{\mu \nu}$ denotes the use of $Q_{\mu \nu}$  and $\sigma_{1 2}^{\mu \rightarrow \nu}(E)$ in Eq. (\ref{eq:R-Formal}).  Inserting Eq.~(\ref{eq:Ratio of Reaction Rates munu}) into Eq.~(\ref{sigmac-star}) and using the fact that $Q_{\mu \nu} = Q - E_{3 \nu} + E_{1 \mu}$,  we can write
\begin{eqnarray}
\langle \sigma c \rangle_{\gamma 3}^* &=&  \bigl(1+R \bigr) {\langle\sigma v\rangle_{12}^*}  \nonumber \\
&\times & \biggl( \frac{ \sqrt{2 \pi}}{8 \zeta(3)} \biggr) 
 \left(\frac{\mu c^{2}}{kT}\right)^{3/2} 
  \frac{G_{1}G_{2}}{G_{3} (1 + \delta_{12})}e^{-Q/kT}~~,
\label{eq:thermal}
\end{eqnarray}
where  $R$ represents an  average correction factor among all thermally populated states.   As demonstrated below, $R$ is nearly independent of the detailed nuclear structure.  Hence, we can simply utilize the ground-state $Q$-value as a representative average over the distribution of $Q$-values among the thermally populated states.   Also,  the spin factors above are now replaced by the relevant nuclear partition functions $G_{i}$ :
\begin{equation}
G_i =\sum_{\alpha}g_{\alpha}e^{-E_{\alpha}/kT}~~,
\end{equation}
where $\alpha$ denotes the individual  states in nucleus $i$.  

Stellar reaction  rate tables are usually listed as functions of temperature $T_9$ in units of $10^9$ K and are given as $\bigl[ N_A \langle \sigma v(T_9) \rangle^* \bigr]$. Thus,  we  can rewrite Eq.~(\ref{eq:thermal}) as
 \begin{eqnarray}
\label{eq:thermalT9}
\lambda_{\gamma 3} &= & \bigl(1+R \bigr)  \bigl[N_A \langle\sigma v(T_9) \rangle^* \bigr]_{12}   \\
 \times 9.8685 \times 10^9 && ({\hat \mu T_9})^{3/2} \frac{G_{1}G_{2}}{G_{3} (1 + \delta_{12})} e^{-11.605Q/T_9}~,\nonumber
\end{eqnarray}
where now $N_A$ is Avagadro's number so that $\bigl[ N_A \langle \sigma v(T_9) \rangle^* \bigr]$ is in units of cm$^3$ mol$^{-1}$ s$^{-1}$, $Q$ is in units of MeV and $\hat \mu$ is the reduced mass in atomic mass units.
 
  Equations (\ref{eq:thermal}) and  (\ref{eq:thermalT9}) are in a convenient form because in the limit of $R \rightarrow 0$, they reduce to usual photodisintegration rates available from various compilations (e.g.,  \cite{fow67,fow75}; \cite{Holmes76}; \cite{Woosley78}; \cite{CF88}; NACRE, \cite{NACRE}; TALYS, \cite{Goriely08}; NONSMOKER, \cite{Rauscher00} or REACLIB, \cite{REACLIB}).  The combined factors multiplying  $ {\langle\sigma v\rangle_{12}^*} $ in Eq.~(\ref{eq:thermalT9}) are usually referred to  as the "reverse ratio" as this factor gives the reverse reaction rate in terms of the forward rate.  In this work, we show that there is a simple  correction $(1 + R)$ to this reverse ratio due to the difference between Planckian and Maxwell-Boltzmann statistics.   For most of the remainder of this manuscript, our goal will be to derive  a simple analytic form for $R$ for ease in correcting existing tabularized reverse reaction rates.   We will also derive simple analytic approximations to clarify the essential physics of this correction and summarize examples  of which astrophysical conditions may be most affected by these correction factors.
  
\subsection{Nuclear Statistical Equilibrium}

Before leaving this discussion, however,  it is worth emphasizing again that the above rate does not imply equilibrium, but only detailed balance and a thermal population of photons and nuclear excited states.   Nevertheless, the situation of equilibrium between capture and photodissociation frequently occurs in astrophysical environments and is referred to as  nuclear statistical equilibrium (NSE).  It is of note that the conditions of NSE are also modified from the usual Saha equation by the above quantum corrections.  Moreover, in conditions of NSE, one  sometimes synthesizes nuclei for which   $Q/kT \sim 1$  and the corrections can become larger.  Examples of this include the formation of nuclei near the proton drip line in the hot hydrogen burning $rp$-process,  or the synthesis of nuclei near the neutron drip line in the neutron-capture $r$-process, as discussed below.  

To see the revised conditions of NSE consider 
the evolution of a nucleus undergoing rapid  particle captures and photodissociation.  This can be written as
\begin{equation}
\frac{dn_1}{dt} = -n_1 n_2  {\langle\sigma v\rangle^*_{12}} + n_3 n_\gamma \langle\sigma c\rangle^*_{\gamma3}~~.
\end{equation}
The equilibrium condition $(dn_1/dt) = 0$  therefore demands that
\begin{equation}
\frac{n_1 n_2}{n_3} =  \frac{n_\gamma  \langle\sigma c\rangle^*_{\gamma3}} {\langle\sigma v\rangle^*_{12}} = \bigl(1+R\bigr)  \biggl(\frac{2 \pi \mu kT}{ h^2}\biggr)^{3/2} 
\frac{G_{1}G_{2}}{G_{3} (1 + \delta_{12})} e^{-Q/kT}~,
\label{sahaeq}
\end{equation}
or in terms of mass fractions and temperature, it is more convenient for stellar models to write
\begin{equation}
\frac{X_1 X_2}{X_3} = 9.8685 \times 10^9 \bigl(1+R\bigr)  \frac{T_9^{3/2} \hat \mu^{5/2}}{\rho}
\frac{G_{1}G_{2}}{G_{3} (1 + \delta_{12})} e^{-11.605Q/T_9}~.
\label{sahaeqT9}
\end{equation}

In the limit that $R \rightarrow 0$,  Eqs.~(\ref{sahaeq}) and (\ref{sahaeqT9}) represent  the usual nuclear Saha equation  \citep{saha21} of  statistical equilibrium which also invokes the Maxwellian approximation given in Eq.~(\ref{approx}) either directly or indirectly in its derivation (cf.~\cite{cla68,Iliadis07}).  The deviation of NSE due to quantum statistics may impact the evolution of explosive nucleosynthesis environments for which one can encounter nuclei with small photodissociation thresholds, e.g., near the neutron or proton drip lines.  To the extent that such nuclei are beta-decay waiting points, for example, the altered statistics  will affect the timescale for the build up of abundances.  Another possible application of the corrections deduced here is for the ionization equilibrium of atomic or molecular species with a low ionization potential in stellar atmospheres.  However, we will not consider that case further here.

\section{Evaluation of the Correction Factor $R$}

It is worth noting that quantum effects always tend to speed up photodisintegration rates because the Planck distribution  places many more  photons at low energy than a Maxwell-Boltzmann distribution of the same temperature or energy density. In other words, $R$ is always  positive definite.  Hence, even though it is often small, it is worth including.
This correction could of course always be evaluated by direct numerical integration of Eq.~(\ref{eq:R-Formal}) or Eq.~(\ref{eq:Inverse Reaction Rate 2}).  In a large network calculation with evolving temperature, however, the repeated numerical integrations would slow the computation time, moreover, it is a tedious task to assemble all of the relevant cross section data.   

Nevertheless, the advantage of introducing the $R$ in Eq.~(\ref{eq:thermal}) is that it gives the quantum correction as a small \emph{fraction} of the classical Maxwellian result.  Hence, for implementation in large networks,   an analytic approximation to the exact numerical integration for $R$ is both adequate and desirable.  Moreover, we show that an accurate analytic correction is readily available based upon the input from existing reaction rate tables (either in analytic or tabularized form).  We also show that $R$ is nearly independent of the nuclear cross sections, and to leading order only depends upon the $Q$-value and the Gamow and/or resonance energy.   

The key to  evaluating  $R$ is to perform a  geometric series expansion of the numerator of Eq. (\ref{eq:R-Formal}), i.e., 
\begin{equation}\label{eq: Full Expansion}
\frac{1}{e^{x}-1}=\sum_{n=1}^{\infty}e^{-nx}.
\end{equation}
Note  that terminating this series after the first term corresponds to the usual Maxwell-Boltzmann approximation in which case one obtains $R=0$ as expected. Using the full expansion, however, leads to
\begin{equation}
R=\sum_{n=2}^{\infty}r_{n}e^{-\left(n-1\right)Q/kT}~,
\label{eq:R}
\end{equation}
where the coefficients $r_{n}$ (for $n\geq2$) are given by
\begin{equation}
r_{n}=\frac{\int_{0}^{\infty}\sigma^*\left(E\right)e^{-nE/kT}EdE}{\int_{0}^{\infty}\sigma^*\left(E\right)e^{-E/kT}EdE}.
\label{eq:rn}
\end{equation}

Eq. (\ref{eq:R}) gives $R$ as a geometric series in powers of $e^{-Q/kT}$ which  is usually a small quantity.  The coefficients $r_{n}$ are also less than unity and they rapidly decrease with increasing $n$. Hence,  the series given in Eq. (\ref{eq:R}) is a rapidly convergent  one.   However, the explicit determination of the $r_n$ requires some attention to the energy dependence of the cross sections.  Nevertheless, this task is greatly simplified when the forward thermonuclear reaction rates as a function of temperature $\langle\sigma v(kT)\rangle$ have  already been compiled.  Combining the expression for the reaction rate in Eq.~(\ref{eq:Forward Reaction Rate}) with Eq.~(\ref{eq:rn}), 
the expansion coefficients become
\begin{equation}
r_n = \frac{1}{n^{3/2}} \frac{\langle\sigma v(kT/n)\rangle^*}{\langle\sigma v(kT)\rangle^*}~~.
\end{equation}
This immediately gives the correction factors in terms of the tabulated rates,
\begin{equation}
R=\sum_{n=2}^{\infty}\frac{1}{n^{3/2}} \frac{\langle\sigma v(kT/n)\rangle^*}{\langle\sigma v(kT)\rangle^*}e^{-\left(n-1\right)Q/kT}~~,
\label{eq:RkT}
\end{equation}
or converting to units of $T_9$ and noting that reaction rate compilations are in terms of $[N_A \langle \sigma v(T_9) \rangle^* ]$ we have
\begin{equation}
R=\sum_{n=2}^{\infty}\frac{1}{n^{3/2}}  \frac{[N_A \langle\sigma v(T_9/n)\rangle^*  ]}{[N_A \langle\sigma v(T_9)\rangle^* ]}e^{-11.605\left (n-1\right)Q/T_9}~.
\label{eq:RT9}
\end{equation}
  
  These expressions for the reverse rate correction factor $R$ make clear the physics of the correction factor.  The factors become a sum of the correction factors in terms of thermonuclear averages over decreasing temperatures, $kT \rightarrow kT/n$.  These terms achieve the task of increasing the photodissociation rate due to the fact that  the Planck distribution includes many more low energy photons as $(E + Q )/kT \rightarrow 0$, which is what the $[\exp{\{(E+Q)/kT\}} -1]$ denominator in  Eq.~(\ref{eq:Inverse Reaction Rate 2}) achieves.  
  
  Eqs. (\ref{eq:RkT}) and (\ref{eq:RT9}) are the key equations for this paper.  Note that $\langle\sigma v(kT)\rangle $ is a rapidly decreasing function as the temperature decreases, as well as the $n^{-3/2}$ pre-factor and the $\exp{[-\left(n-1\right)Q/kT]}$. These three conditions  guarantee that this is a well behaved, rapidly decreasing, convergent series even as $Q \rightarrow 0$.  
In what follows, we show some illustrations of the magnitude of these corrections and also derive some alternative analytic forms to illustrate the basic structure of these corrections in more detail.   Indeed, we show that in practice, only one or two terms are needed in the series and the correction factors  are largely independent of the underlying nuclear structure.  As a caveat to the reader, however,  we note that the analytic approximations derived below do not include the stellar enhancement factors, and as such should be used with caution in a real astrophysical plasma.

\subsection{Non-resonant Charged-particle Capture Reactions}

When the projectile particle is charged, the capture reaction must
 tunnel through the Coulomb barrier at low energy. To account for this, the cross section can be factored into the following form \citep{fow67}:
\begin{equation}
\sigma \left(E\right)=S\left(E\right)\frac{1}{E}e^{-\sqrt{\frac{E_{G}}{E}}}.
\label{eq:Cross Section}
\end{equation}
Here, $S\left(E\right)$ is called the astrophysical $S$-factor and $E_{G}$ is the Gamow energy which characterizes the penetrability:
\begin{eqnarray}
E_{G}&=&\frac{1}{2}\mu\left(\frac{2\pi Z_{1}Z_{2}e^{2}}{\hbar}\right)^{2} \\
&&  =0.97907~ W~{\rm (MeV)}~, \nonumber
\label{eq:Gamov Energy}
\end{eqnarray}
where
\begin{equation}
W = Z_{1}^2Z_{2}^2 \hat \mu~~,
\end{equation}
and $\hat \mu$ is in atomic mass units.

The $S$ factor  contains
 information about the detailed nuclear interaction.
Away from resonances, the astrophysical $S(E)$ factor is a slowly varying function, and at low temperatures the integrand is dominated by a small region known as the Gamow window \citep{fow67} located at an energy $E_0$ as defined below.  As such, to the desired accuracy  it can be replaced with an average effective value,  $S_{\mbox{\footnotesize eff}}(E_0).$  Therefore, at low temperature $S(E)$ will  cancel in the ratio given in Eq. (\ref{eq:rn}).  However, for  higher temperatures  the variation of the $S$-factor with energy over the Gamow window can become relevant.  Hence, following \cite{fow67} we write
\begin{equation}
S(E) = S(0)\biggl[1  + \frac{\dot S(0)}{S(0)} E + \frac{1}{2}  \frac{\ddot S(0)}{S(0)} E^2 + \dots\biggr]~,
\end{equation}
where the dot denotes derivative with respect to energy. 

Inserting this cross section into Eq. (\ref{eq:rn})  the correction coefficient $r_{n}$ becomes 
\begin{equation}
r_{n}=\frac{F\left(n\frac{E_{G}}{kT}\right)}{F\left(\frac{E_{G}}{kT}\right)},
\label{eq:rn3}
\end{equation}
where the function $F$ is defined as follows: 
\begin{equation}
F\left({E_{G}}/{kT} \right)=\int_{0}^{\infty}S(E) e^{-\sqrt{{E_G}/{E}}}e^{-E/kT}dE.
\label{eq:Integral}
\end{equation}

This is a familiar integral in nuclear astrophysics \citep{fow67}.  Eq.~(\ref{eq:Integral})  corresponds to a product of the barrier penetrability times a Maxwellian distribution.  It is strongly maximized in the Gamow window and well approximated \citep*{fow67,cla68,Iliadis07} as a Gaussian integral near the maximum of the integrand. Hence, we write
\begin{equation}
F({E_{G}}/{kT}) = \biggl(\frac{\pi}{4}\biggr)^{1/2}\Delta S_{\mbox{\footnotesize eff}}(kT) ~e^{-{3E_0}/{kT}}~~,
\label{eq:fg}
\end{equation}
where  
\begin{eqnarray}
E_0 &=& (\sqrt{E_G}kT/2)^{2/3} \nonumber \\
&& = 0.12204 ~W^{1/3} T_9^{2/3}
\label{E0}
\end{eqnarray}
is the peak of the Gamow window and
\begin{eqnarray}
\Delta &=& 4 \sqrt{(E_0 kT/3)} \nonumber \\
&=& 0.23682 ~W^{1/6} T_9^{5/6}  ~{\rm MeV}
\label{Delta}
\end{eqnarray}
is the peak width. The effective $S$-factor for charged-particle reactions is given by \cite{fow67}:
 \begin{eqnarray}
S_{\mbox{\footnotesize eff}}(kT)  &= & S(0)\biggl[ 1 + \frac{5}{12 \tau} +  \frac{\dot S(0)}{S} \bigl( E_0 + ({35}/{36})kT \bigr) \nonumber \\
 +&& \frac{1}{2} \frac{\ddot S(0)}{S} \bigl( E_0^2 + ({89}/{36}) E_0 kT\bigr) \biggr]~~, 
\label{seffkt}
\end{eqnarray}
where $\tau \equiv 3 E_0/(kT)$.  Converting to temperature, this becomes
\begin{eqnarray}
S_{\mbox{\footnotesize eff}}(T_9)  &= & S(0)\biggl[ 1 + 0.09807T_9^{1/3} W^{-1/3} + 0.1220  \frac{\dot S(0)}{S} W^{1/3} T_9^{2/3}\nonumber \\
+ && 0.08378  \frac{\dot S(0)}{S} T_9 + 7.447\times 10^{-3} \frac{\ddot S(0)}{S} W^{2/3} T_9^{4/3}  \nonumber \\
+ && 0.01300 \frac{\ddot S(0)}{S} W^{1/3} T_9^{5/3} \biggr]~~{\rm MeV-barn}~.
\label{sefft9}
\end{eqnarray}
The evaluation of the remaining terms in the series can be done simply by making the replacement $kT \rightarrow kT/n$  in Eq.~(\ref{eq:fg}).  

Now collecting all terms, the  coefficient  $r_n$ becomes
\begin{eqnarray}
\label{eq:rn3analy}
r_n &=& \frac{S_{\mbox{\footnotesize eff}}(kT/n) }{n^{5/6} S_{\mbox{\footnotesize eff}}(kT) }e^{-\bigl[ 3\bigl({E_G}/{4 kT}\bigr)^{1/3}\bigl(n^{1/3} - 1\bigr)\bigr]} \\
 =  &&\frac{S_{\mbox{\footnotesize eff}}(T_9/n) }{n^{5/6} S_{\mbox{\footnotesize eff}}(T_9) }e^{-\bigl[ 4.2487(W/T_9)^{1/3}\bigl(n^{1/3} - 1\bigr)\bigr]} ~~,\nonumber
\end{eqnarray} 
 which immediately leads to the desired analytic expression for $R$:
\begin{eqnarray}
\label{eq:RwS}
R &=& \sum_{n=2}^{\infty} \frac{S_{\mbox{\footnotesize eff}}(kT/n) }{n^{5/6} S_{\mbox{\footnotesize eff}}(kT) }e^{-\bigl[{(n-1)Q}/{kT} + 3\bigl({E_G}/{4 kT}\bigr)^{1/3}\bigl(n^{1/3} - 1\bigr)\bigr]} \\
&=& \sum_{n=2}^{\infty} \frac{S_{\mbox{\footnotesize eff}}(T_9/n) }{n^{5/6} S_{\mbox{\footnotesize eff}}(T_9) }e^{-\bigl[{11.605(n-1)Q}/{T_9} +  4.2487(W/T_9)^{1/3}\bigl(n^{1/3} - 1\bigr)\bigr]} ~~,\nonumber
\end{eqnarray}
where $Q$ is in MeV in the latter equation.  Note that the dependence on $S(0)$ cancels in the $r_n$ and the correction factor $R$.
Also note that as the ratio of the $S_{\mbox{\footnotesize eff}}$ factors approaches unity, the correction factor only depends upon the Gamow energy and the reaction $Q$-value. Hence, for a slowly-varying $S$-factor we have
\begin{eqnarray}
\label{eq:rnApproximate}
r_n& \approx  & \frac{1 }{n^{5/6}  }e^{-\bigl[ 3\bigl({E_G}/{4 kT}\bigr)^{1/3}\bigl(n^{1/3} - 1\bigr)\bigr]} \\
  &=&\frac{1 }{n^{5/6}  }e^{-\bigl[ 4.2487(W/T_9)^{1/3}\bigl(n^{1/3} - 1\bigr)\bigr]} \nonumber
\end{eqnarray}
 and
\begin{eqnarray}
\label{eq:RApproximate}
R &=& \sum_{n=2}^{\infty} \frac{1 }{n^{5/6} } e^{-\bigl[{(n-1)Q}/{kT} + 3\bigl({E_G}/{4 kT}\bigr)^{1/3}\bigl(n^{1/3} - 1\bigr)\bigr]} \\
&=& \sum_{n=2}^{\infty} \frac{1 }{n^{5/6}  }e^{-\bigl[{11.605(n-1)Q}/{T_9} +  4.2487(W/T_9)^{1/3}\bigl(n^{1/3} - 1\bigr)\bigr]} ~~.\nonumber
\end{eqnarray}

\begin{figure}
\begin{center}
\vspace{3mm}
\includegraphics[scale=0.6]{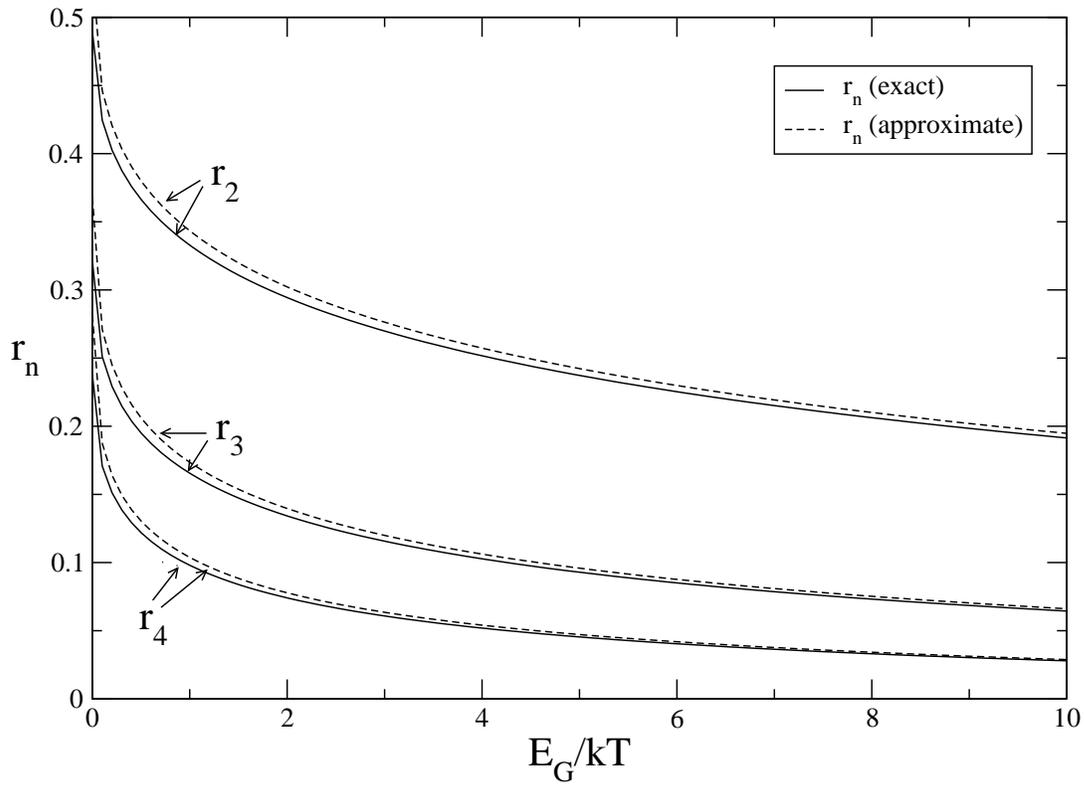}
\end{center}
\caption{First few correction coefficients for charged-particle reactions as functions of $E_{G}/kT$ for the case of a constant $S$-factor.  The dashed lines show the approximate values based upon Eq. (\ref{eq:rn3analy}) and the solid lines show an exact numerical integration of Eq.~(\ref{eq:rn}).}
\label{fig:Coefficients}
\end{figure}

In practice, this is a rapidly converging series requiring at most the first few terms for sufficient accuracy.  
To illustrate the rate of convergence, the first few $r_n$ coefficients of Eq.~(\ref{eq:rnApproximate}) 
are plotted as a function of  $E_{G}/kT$ in Figure \ref{fig:Coefficients}.  It is clearly seen in this figure that these coefficients  become small as $n$ increases and that the terms in the series are nearly of the same functional form.  Hence, if speed is desired,  an adequate approximation can usually be obtained by retaining only the first term in the series for $R$.  For illustration the dashed lines in Figure \ref{fig:Coefficients} show a comparison of the exact integration of Eq.~(\ref{eq:rn3}) with the analytic expansion (Eq.~(\ref{eq:rnApproximate})) in the case of a slowly varying $S$-factor.  Clearly, the analytic series is adequate for the $r_n$. The slight difference between these two sets of curves relates to the fact that the Gamow window is not exactly Gaussian even in the case of a constant $S$-factor.   This, for example, is the reason that for a constant $S$ factor as $E_G \rightarrow 0$, the exact expression in Eq.~(\ref{eq:rn3}) goes to $1/n$ as can be seen from Eq.~(\ref{eq:Integral}), whereas the analytic approximation expansion goes to $1/n^{5/6}$.  

\begin{figure}
\begin{center}
\includegraphics[scale=0.5]{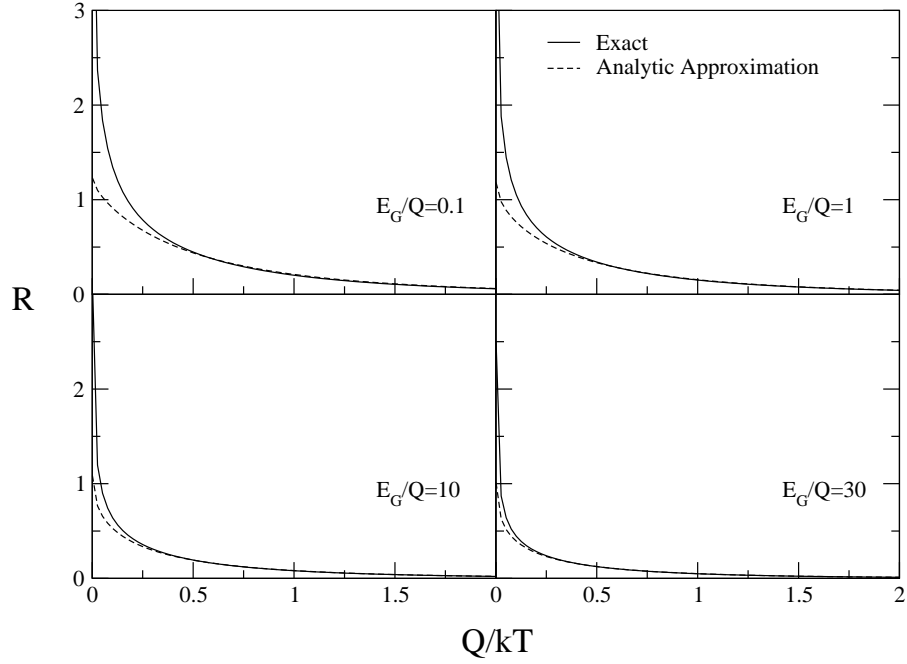}
\end{center}
\caption{Reverse rate correction factor $R$ for charged-particle reactions as a function of $Q/kT$ for various  values of $E_G/Q$. 
The solid lines in this figure are from a  numerical integration  of Eq.~(\ref{eq:R-Formal}) for the case of  a constant $S$ factor. The dashed lines show the result of the series expansion in Eq. (\ref{eq:RApproximate}) with only the first three terms included. }
\label{fig:nonreschar}
\end{figure}

Figure \ref{fig:nonreschar} illustrates the charged-particle reverse-reaction correction factor $R$ from Eq.~(\ref{eq:RApproximate}) as a function of $Q/kT$ for several values of the dimensionless quantity $E_G/Q$. 
The solid line is based upon the numerical integration of Eq. (\ref{eq:R-Formal}) with a constant $S(E)$ factor while the dashed line is based upon the first three terms of the approximate series given in Eq. (\ref{eq:RApproximate}).
Even though the $r_n$ remain large up to  $E_G/kT> 10$,  the total  correction $R$ is only  $ ^>_\sim 1$\% for reactions for which $Q/kT ^<_\sim 1$.  From this figure it is also clear that retaining only the first three terms is an excellent approximation down to $Q/kT \sim 0.4$.  Below that, however one should probably include more terms in the expansion depending upon the $E_G/Q$ ratio.  We have found, however, that one never needs more than about six terms in the series even in the limit that $Q \rightarrow 0$.

\begin{figure}
\begin{center}
  \subfigure[]{
  \includegraphics[scale=0.3]{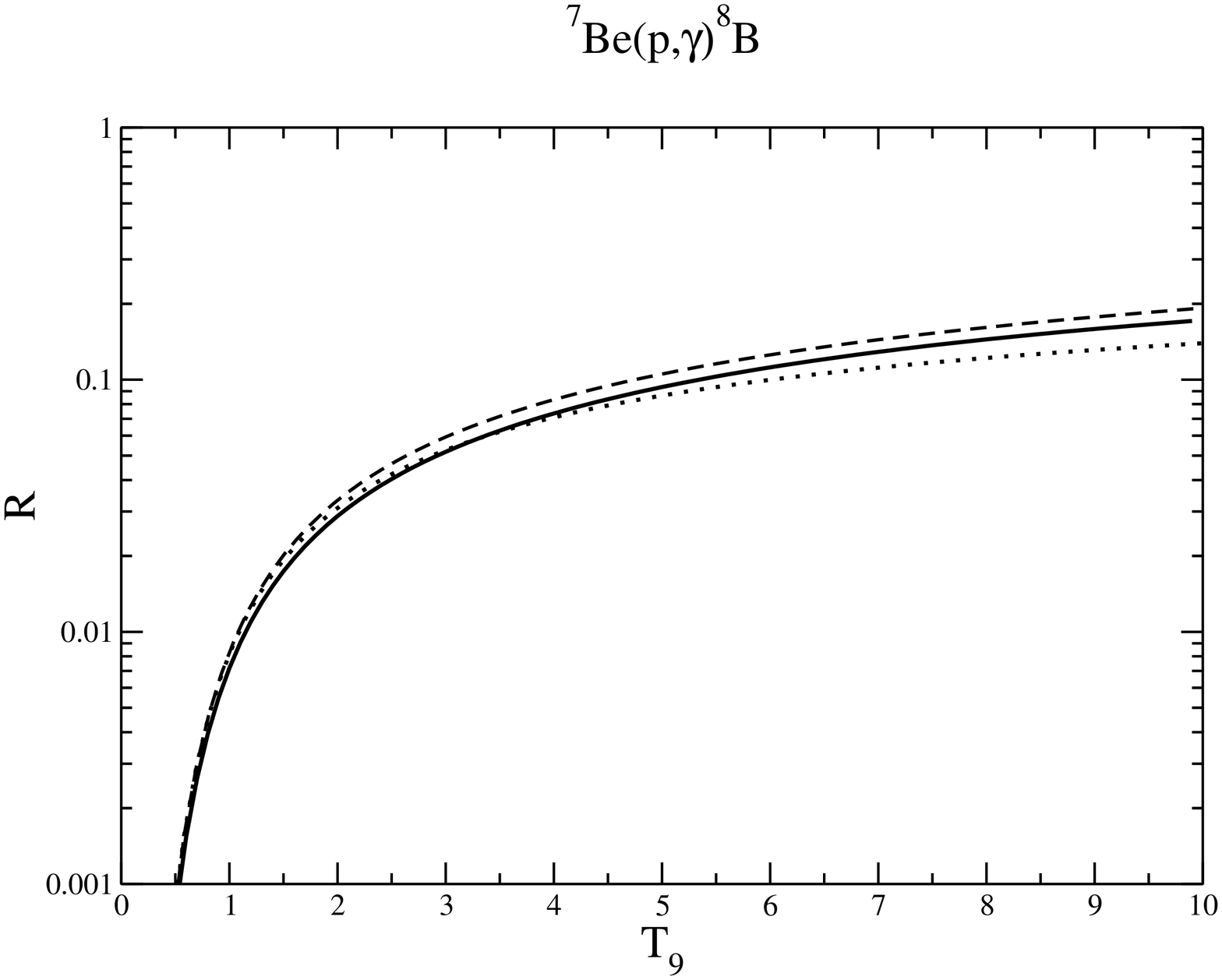}
  }
  \subfigure[]{
  \includegraphics[scale=0.3]{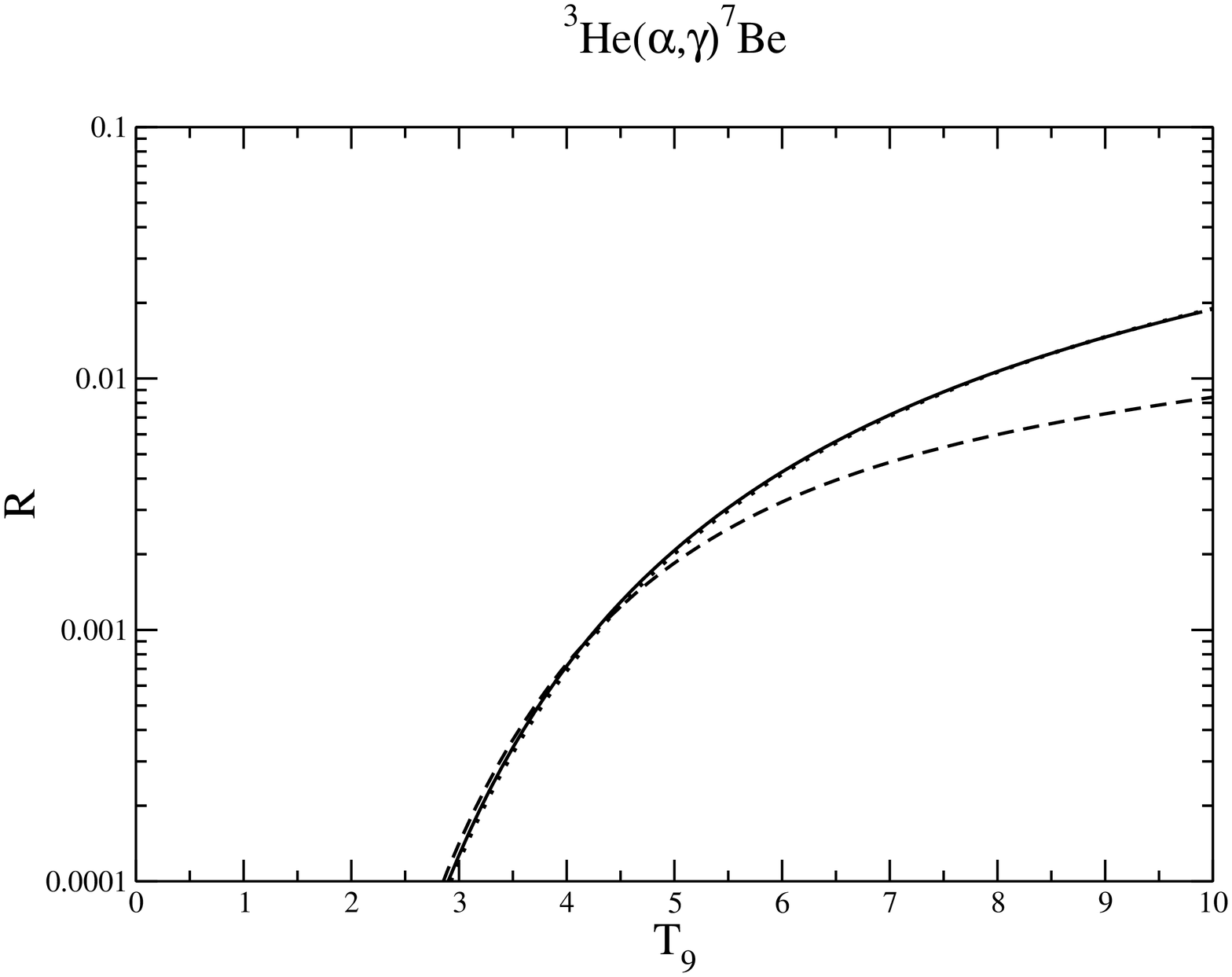}
  }
\end{center}
\caption{Reverse reaction rate correction factor $R$ for the hydrogen-burning charged-particle reactions $^7$Be$(p,\gamma)^8$B ($Q = 0.135$ MeV)  and  $^3$He$(\alpha,\gamma)^7$Be  ($Q = 1.587$ MeV)  as a function of $T_9$.  These plots were generated from Eq.~(\ref{eq:RT9}) using the REACLIB compilation \citep{REACLIB} for $[N_A \langle \sigma v\rangle(T_9)]$ and reaction $Q$-values.  On each plot, the solid line shows the application of Eq.~(\ref{eq:RT9}), the dashed line shows the application of Eqs.~(\ref{eq:RwS}), while the dotted line shows the results obtained from keeping only the first term in the series in Eq.~(\ref{eq:RApproximate}). For the $^7$Be$(p,\gamma)^8$B a constant $S$-factor was assumed, while the dashed line for  the$^3$He$(\alpha,\gamma)^7$Be  reverse reaction is based upon  $\left(  \dot{S}(0)/S(0), \ddot{S}(0)/2S(0) \right)=\left(-0.953 {\rm~MeV}^{-1}, 0.835 {\rm~ MeV}^{-2}\right)$  from \cite{CF88}.  
}
\label{fig:Correction for H-burn}
\end{figure}

As illustrations of corrections to the reverse ratio for  an evaluated table, the plots on Figure \ref{fig:Correction for H-burn}  show  the $R$ factors  for the hydrogen-burning  charged-particle reactions $^7$Be$(p,\gamma)^8$B ($Q = 0.135$ MeV) and  $^3$He$(\alpha,\gamma)^7$Be  ($Q = 1.587$ MeV)  as a function of $T_9$.   These plots were generated from Eq.~(\ref{eq:RT9}) using the REACLIB compilation \citep{REACLIB} for $[N_A \langle \sigma v\rangle(kT)]$ and reaction $Q$-values.  Although most hydrogen burning takes place at relatively low temperature ($T_9 < 1$), evaluated thermonuclear reaction rates are usually tabularized up to $T_9 = 10$. Hence, we consider the same range here.  

On each plot, the solid line shows the application of Eq.~(\ref{eq:RT9}), the dashed line shows the application of Eqs.~(\ref{eq:RwS}), while the dotted line shows the results obtained from keeping only the first term in the series in Eq.~(\ref{eq:RApproximate}). Surprisingly, the correction for the reverse $^3$He$(\alpha,\gamma)^7$Be reaction based upon the REACLIB evaluation  is almost indistinguishable from the approximation based upon only the first term in the series with a constant $S$-factor.  This is due to the fact that the $S$ factor in the compilation is taken to be constant at high energies and that  $Q/kT > 1$ for the plotted range of $T_9$. These figures  show that  the corrections are generally quite small and adequately represented by only the first term in Eq.~(\ref{eq:RApproximate}).  

\placefigure{fig:Coefficients}

\subsection{Non-resonant Neutron-capture Reactions}

\begin{figure}
\begin{center}
\includegraphics[scale=0.5]{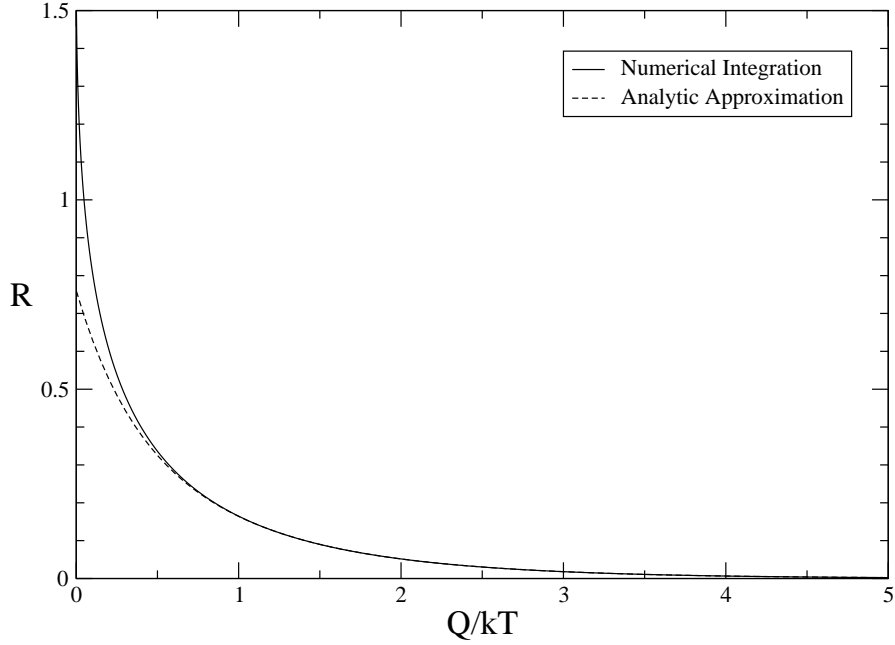}
\end{center}
\caption{Reverse reaction rate correction factor $R$ as a function of  $Q/kT$ for nonresonant neutron-capture reactions with the assumption of a slowly varying ${\cal S}$ factor.  The solid line is from an  exact numerical integration of Eq.~(\ref{eq:R-Formal}).  The dashed line is from the analytic expression given in Eq. (\ref{nonrescorr}) truncated after the first three terms.}
\label{fig:R}
\end{figure}

At low energies where the de Broglie wavelength
of the neutron is much larger than the radius of the target
nucleus, the non-resonant neutron-capture cross section is proportional to $1/v$. Hence,  
\begin{equation}
\sigma_{12}(E)\sqrt{E}\approx\mbox{constant}~~.
\label{eq:Low Energies}
\end{equation}
 At high energies, however, deviations from Eq. (\ref{eq:Low Energies}) occur. Such deviations can be described in terms of a 
Maclaurin series in $E^{1/2}$  \citep{fow67} which roughly accounts for the contribution of higher partial waves to the cross section:
\begin{equation}
\sigma_{12} v={\cal S}(0)\biggl[ 1 + \frac{\dot {\cal S} (0)}{{\cal S}(0) } E^{1/2} + \frac{1}{2} \frac{\ddot {\cal S} (0)}{{\cal S} (0) } E +\dots \biggr]
\label{Fowler}
\end{equation}
where  the dot denotes  a derivative with respect to $E^{1/2}$. The integration of Eq.~(\ref{Fowler}) over a Maxwellian velocity distribution gives  \citep{fow67}
 \begin{equation}
 \langle \sigma_{12} v \rangle = {\cal S} _{\mbox{\footnotesize eff}}(kT)~~,
 \end{equation} 
 where ${\cal S} _{\mbox{\footnotesize eff}}(kT)$ is defined by
\begin{eqnarray}
{\cal S} _{\mbox{\footnotesize eff}}(kT)  &\equiv& {\cal S} (0) \biggl[ 1 + \frac{2}{\sqrt{\pi}} \frac{\dot {\cal S} (0)}{{\cal S} (0)} (kT)^{1/2} + \frac{3}{4} \frac{\ddot {\cal S} (0)}{{\cal S} (0)} kT + \dots \biggr]  \\
 = {\cal S} (0) \biggl[ 1& + & 0.3312\frac{\dot {\cal S} (0)}{{\cal S}(0) } (T_9)^{1/2} + 0.06463 \frac{\ddot {\cal S} (0)}{{\cal S}(0) } T_9  + \dots \biggr] ~{\rm cm^3 s^{-1} }~.\nonumber
\label{seff-neutron}
 \end{eqnarray}
 Inserting this into Eq.~(\ref{eq:R}) then immediately gives,
\begin{eqnarray}
\label{R for neutron capture}
R&=& \sum_{n=2}^{\infty}\frac{1}{n^{3/2}}\frac{{\cal S} _{\mbox{\footnotesize eff}}(kT/n)}{{\cal S} _{\mbox{\footnotesize eff}}(kT)}e^{-(n-1)Q/kT }  \\
  = && \sum_{n=2}^{\infty} e^{- 11.605 (n-1)Q/T_9 }  \nonumber \\
 && \times \frac{1}{n^{3/2}} \frac{ \bigl[ 1 + 0.3312\frac{\dot {\cal S} (0)}{{\cal S}(0) } (T_9/n)^{1/2} + 0.06463 \frac{\ddot {\cal S} (0)}{{\cal S} (0)} (T_9/n)  \bigr]  }
 {\bigl[ 1 + 0.3312\frac{\dot {\cal S} (0)}{{\cal S}(0) } (T_9)^{1/2} + 0.06463 \frac{\ddot {\cal S} (0)}{{\cal S}(0) } T_9 \bigr]  }
 \nonumber~~,
\end{eqnarray}
in which the leading ${\cal S} (0)$ factor obviously cancels. In the astrophysical environments  of most relevance for the present application
one can often ignore the derivatives of ${\cal S} (0)$.  We are also mainly concerned with nuclei and environments with  $Q/kT^>_\sim 1$  for which Eq.~(\ref{R for neutron capture}) quickly converges.  Hence, retaining only the first few terms in the expansion we can write a sufficiently accurate correction as
\begin{eqnarray}
R& \approx & \frac{e^{- Q/kT}}{2^{3/2}} + \frac{e^{- 2Q/kT}}{3^{3/2}} + \frac{e^{- 3Q/kT}}{4^{3/2}} + \cdot \cdot \cdot \nonumber \\
\approx && \frac{e^{- 11.605Q/T_9}}{2^{3/2}}  + \frac{e^{- 22.210Q/T_9}}{3^{3/2}}  + \frac{e^{- 34.815Q/T_9}}{4^{3/2}}  + \cdot \cdot \cdot ~~.
\label{nonrescorr}
\end{eqnarray}
Figure \ref{fig:R}  shows the reverse rate correction factor for nonresonant neutron capture as a function of $Q/kT$.  The solid line is for an  exact numerical integration of Eq.~(\ref{eq:R-Formal}).  The dashed line is from the analytic expression given in Eq. (\ref{nonrescorr}) truncated after the first three terms.  The first few terms in Eq.~(\ref{nonrescorr}) are an adequate approximation until $Q/kT ^<_\sim 0.2$, below which one should include more terms.   

\subsection{Resonant Capture  Reactions}

Often one encounters charged-particle and neutron-capture reactions in which the thermonuclear reaction rates can be dominated by single (or a few) low-lying resonances.  In such cases,  $\sigma_{12}$ in Eq.~(\ref{eq:R-Formal}) is replaced by the Breit-Wigner resonant capture cross section for each resonance,
\begin{equation}
\label{Breit-Wigner}
\sigma_{12}(E) = \pi  \lambdabar^2 \frac{\omega_r \Gamma_i \Gamma_\gamma}{(E-E_r)^2 + (\Gamma_{\mbox{\footnotesize tot}}^2/4)}~~,
\end{equation}
where $\lambdabar$ is the de Broglie wavelength, $\omega_r = g_r (1 + \delta_{1 2})/(g_1 g_2)$ includes spin factors of the reaction, while $\Gamma_i$ is the particle (e.g., proton or neutron) width, $\Gamma_\gamma$ is the width for gamma decay from the resonant state, and $E_r$ is the observed resonance energy.  

It is worthwhile to consider  the limit in which  the total resonance width is small and  dominates the reaction rate.  In that case, in the limit of $\Gamma_{\mbox{\footnotesize tot}} \rightarrow 0$,  we have
\begin{equation}
\frac{1}{\pi} \frac{\Gamma_{\mbox{\footnotesize tot}}/2}{(E-E_r)^2 +( \Gamma_{\mbox{\footnotesize tot}}^2/4)} \rightarrow \delta(E - E_r)~~.
\end{equation}
Inserting this into  Eq.~(\ref{eq:rn}) for $r_n$, the integrals are greatly simplified and reduce to
\begin{equation}
r_n = e^{-(n-1)E_r/kT}~~.
\end{equation}
After summing the series, this leads to the final correction factor for a single resonance of
\begin{eqnarray}
\label{R_Resonant}
R &=& \frac{e^{(E_r+Q)/kT}}{e^{(E_r+Q)/kT}-1}-1  \\
&& = \frac{e^{11.605(E_r+Q)/T_9}}{e^{11.605(E_r+Q)/T_9}-1}-1  ~~, \nonumber
\end{eqnarray}
where, in the second equation the resonance energy and $Q$-value are in units of MeV.  As in the above cases, this correction factor is only  $> 1$\% in reactions for which $(E_r + Q)/kT \sim 1$.  

\begin{figure}
\begin{center}
\includegraphics[scale=0.5]{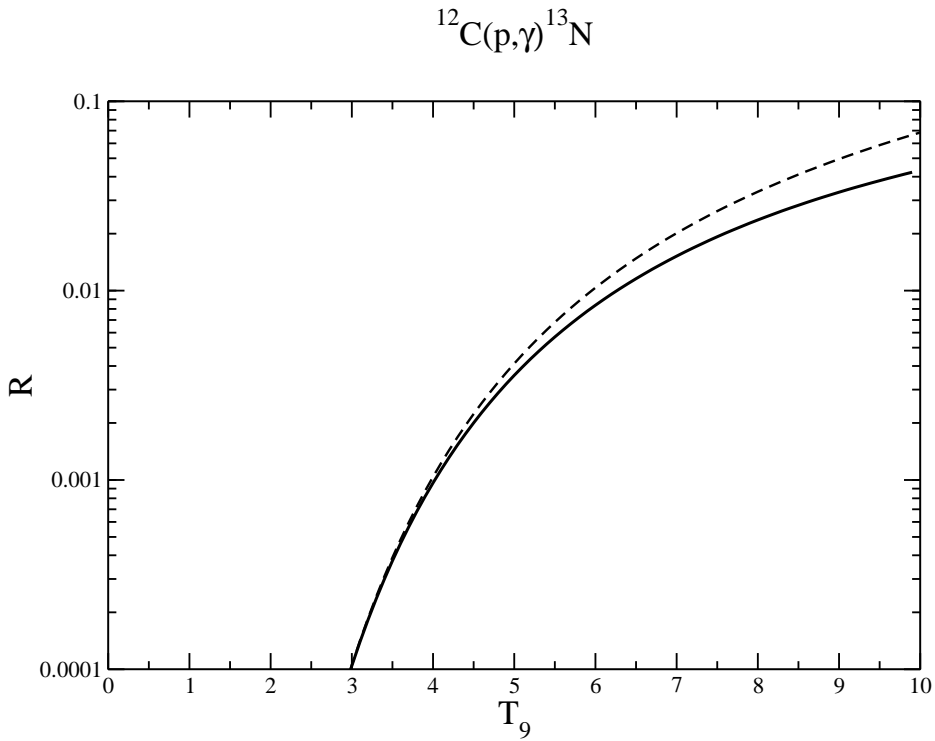}
\end{center}
\caption{Reverse ratio correction factor $R$ for the resonant hydrogen-burning charged-particle reaction $^{12}$C$(p,\gamma)^{13}$N ($Q = 1.944$ MeV)   as a function of $T_9$.  The solid  line was generated from Eq.~(\ref{eq:RT9}) using the REACLIB compilation \citep{REACLIB} for $[N_A \langle \sigma v\rangle(T_9)]$ and reaction $Q$-values.   The dashed line shows the result from application of the simple single resonance correction in Eq.~(\ref{R_Resonant}) for a resonance energy of $E_r = 0.434$ MeV.  
}
\label{R-C12pg}
\end{figure}

As an illustration of  a  resonant reaction, Figure \ref{R-C12pg} shows the reverse rate correction factor for the resonant reaction $^{12}$C$(p,\gamma)^{13}$N ($Q = 1.944$ MeV) as a function of $T_9$.  The solid line on this  plot was generated from Eq.~(\ref{eq:RT9}) using the REACLIB compilation \citep{REACLIB} for $[N_A \langle \sigma v\rangle(T_9)]$ and the reaction $Q$-value, while the dashed line shows the result from an application of the simple single resonance correction in Eq.~(\ref{R_Resonant}).  
Even though this reaction has a second resonance at higher energy, most of the correction factor is accounted for  by the single resonance approximation.   

\section{Applications}
\label{applications}

Having deduced the analytic corrections  it is worthwhile to briefly consider some  illustrations of the practical applications of the above corrections.  From the discussion above it is clear that these correction factors arise as $Q/kT \sim 1$.  For practical applications in astrophysics, this implies $T_9~ ^> _\sim 1$.  We now  consider several  astrophysical environments in which these quantum corrections might appear.  These include the rapid neutron-capture reactions near the neutron drip line during the $r$-process,  the rapid proton capture near the proton drip line during  the $rp$-process, core or explosive oxygen or silicon burning in massive stars,  the $\gamma$-process formation of proton-rich nuclei, and the first few moments of cosmic expansion during the epoch of big bang nucleosynthesis.

\subsection{ $r$-Process}

The $r$-process involves  a sequence of rapid neutron captures in an explosive environment \citep{B2FH,Mathews85}.  It is responsible for the production of about half of the observed abundances of elements heavier than iron.  Although many sites have been proposed for the $r$-process, the neutrino energized wind above the proto-neutron star in core-collapse supernovae remains a favorite \citep{Woosley94}.  Whatever the environment, however, it can be shown that  the solar-system $r$-process abundances are well reproduced by beta-decay flow in a system that is in approximate $(n,\gamma) \leftrightarrows (\gamma,n)$ equilibrium.
Hence, the relative abundances of isotopes of a given element are determined by the revised nuclear Saha equation (\ref{sahaeq})
\begin{equation}
\frac{n(A) }{n(A+1)} =  \bigl(1+R\bigr)\frac{1}{n_n}  \biggl(\frac{2 \pi \mu kT}{ h^2}\biggr)^{3/2} 
\frac{G_{A}G_{n}}{G_{A+1} } e^{-Q_n/kT}~~,
\label{rproc}
\end{equation}
where $Q_n$ is the neutron-capture $Q$-value for isotope $^{A}Z$ (or equivalently the neutron separation energy for the nucleus $^{A+1}Z$) and $n(A)$ represents the abundance of an isotope $^{A}Z$.  This equation defines a sharp peak in abundances for one (or a few) isotopes within an isotopic chain.   The flow of beta decays along these peak isotopes is then known as the $r$-process path.

The location of the $r$-process path peak is roughly identified \citep{B2FH} by the condition that neutron capture ceases to be efficient once  $n(A+1)/n(A)\sim 1$.  Taking the logarithm of Eq.~(\ref{rproc}) and inserting the numerical terms, the $r$-process path can be identified by the following relation:
\begin{eqnarray}
& &\biggl(\frac{Q_n}{kT}\biggr)_{\mbox{\footnotesize path}} = \label{path} 
\\  & & 2.30 \biggl(35.68 + \frac{3}{2}\log{(\frac{kT}{\mbox{\footnotesize MeV}})} -  \log{(\frac{n_n}{\mbox{\footnotesize cm}^{-3}})} + \log{(1 + R)}\biggr)~~.\nonumber
\end{eqnarray}

For a typical $r$-process temperature of $T_9\sim 1 $, the requirement that the $r$-process path reproduce the observed abundance peaks at $A = 80, ~130,$ and $195$ implies that the $r$-process path waiting point in beta flow halt near the neutron closed-shell nuclei $^{80}$Zn, $^{130}$Cd and  $^{195}$Tm.  For a neutron density sufficiently high ($n_n \sim 10^{20}$ cm$^{-3}$) so that the neutron-capture rates exceed the beta-decay rates for these isotopes the peak abundances along the $r$-process path must be for isotopes with $Q_n \sim 3$ MeV, and thus $(Q_n/kT)_{\mbox{\footnotesize path}} \sim 30$.  For this ratio, the correction factors in Eq.~(\ref{nonrescorr}) are negligible. This constraint on $Q_n$, however, concerns the conditions near "freezeout" when the final neutrons are exhausted at the end of the $r$-process.   At this point, the system falls out of NSE and the $r$-process path decays back to the line of stable isotopes.  
 
 From Eq.~(\ref{path}) we deduce that $Q/kT \sim 1$ along the path requires a  neutron density of $\sim 10^{30}$ cm$^{-3}$. Although such conditions do not occur at freezeout, they may occur earlier in the $r$-process.   For example, in the neutrino driven wind models of \cite{Woosley94}, the $r$-process conditions begin with the neutron density at $n_n \approx 10^{27}$  cm$^{-3}$ and a temperature of $T_9 \sim 2$.  The density is also much higher ($> 10^{32}$ cm$^{-3}$) when the  material is first ejected from the proto-neutron star.   Such conditions may also be achieved for an $r$-process which occurs during neutron-star mergers \citep{Freib99,Rosswog99,Rosswog00}.  
 
 As the correction factors become larger, the effect on the $r$-process  may be to slightly increase  the  time it takes for the $r$-process to build up substantial abundances of the  heaviest nuclei.  This is because the faster photodisintegration rates will drive the $r$-process path slightly closer to the line of beta stability.
 
\begin{figure}
\begin{center}
  \subfigure[]{
  \includegraphics[scale=0.3]{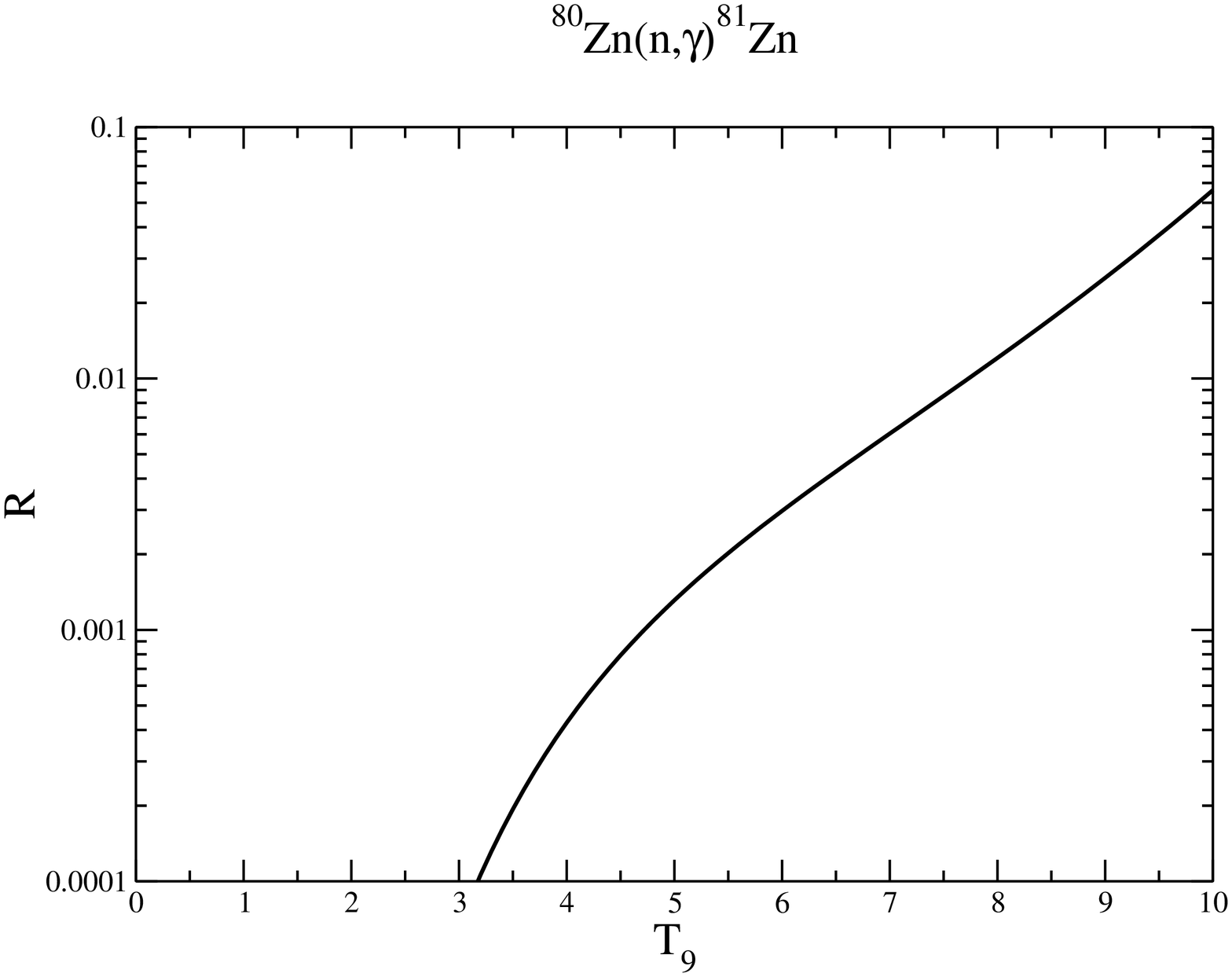}
  }
  \subfigure[]{
  \includegraphics[scale=0.3]{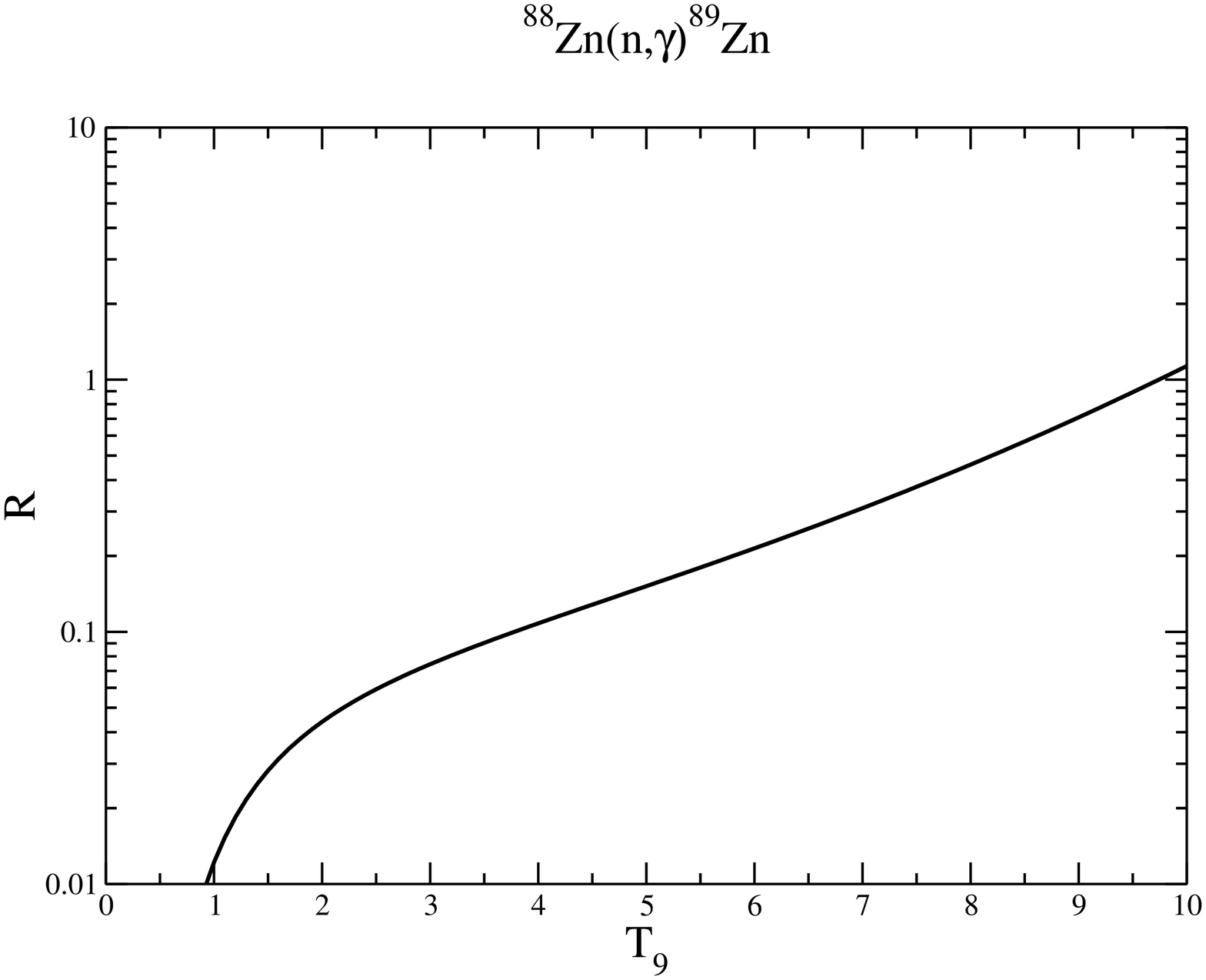}
  }
  
  \subfigure[]{
  \includegraphics[scale=0.3]{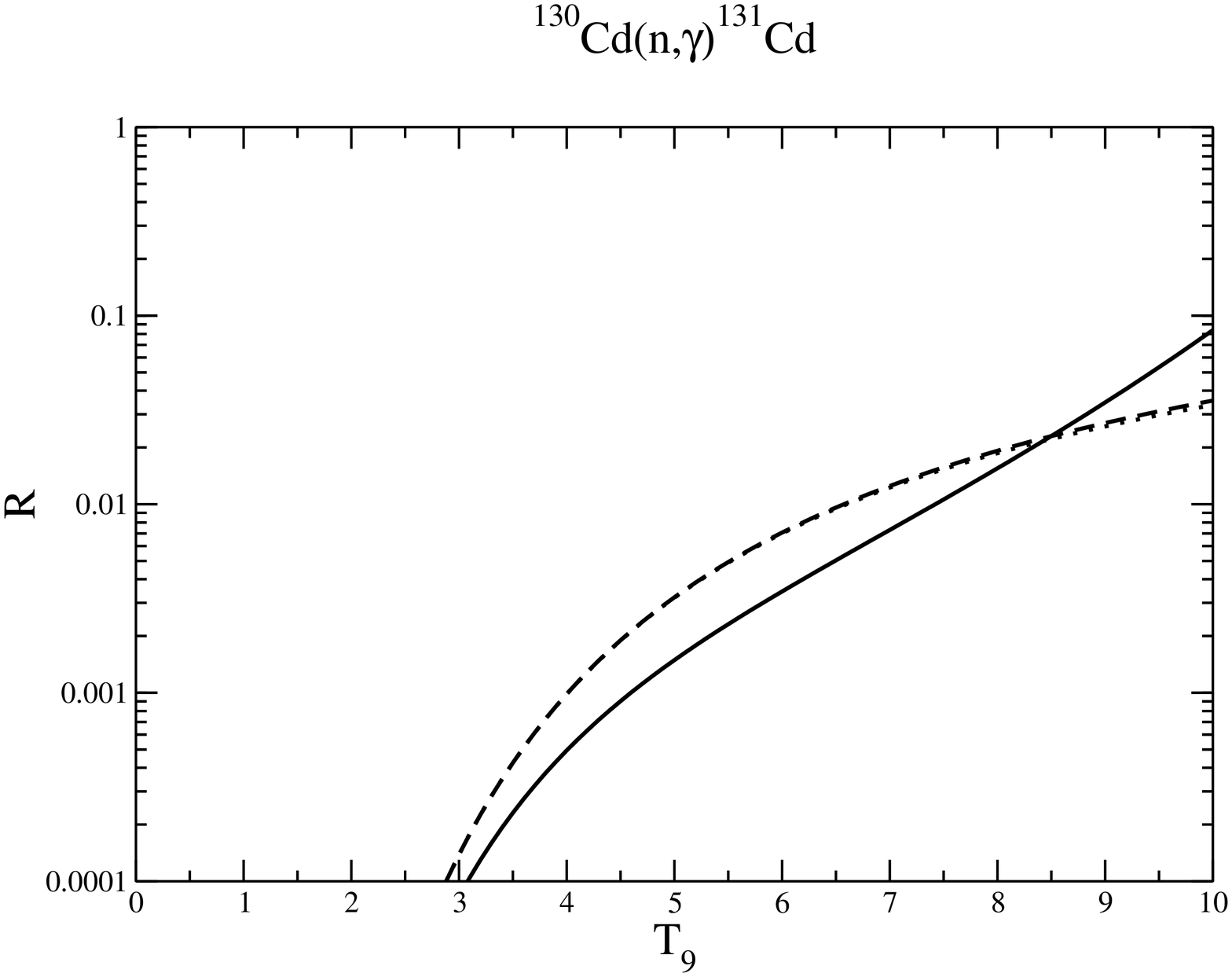}
  }
  \subfigure[]{
  \includegraphics[scale=0.3]{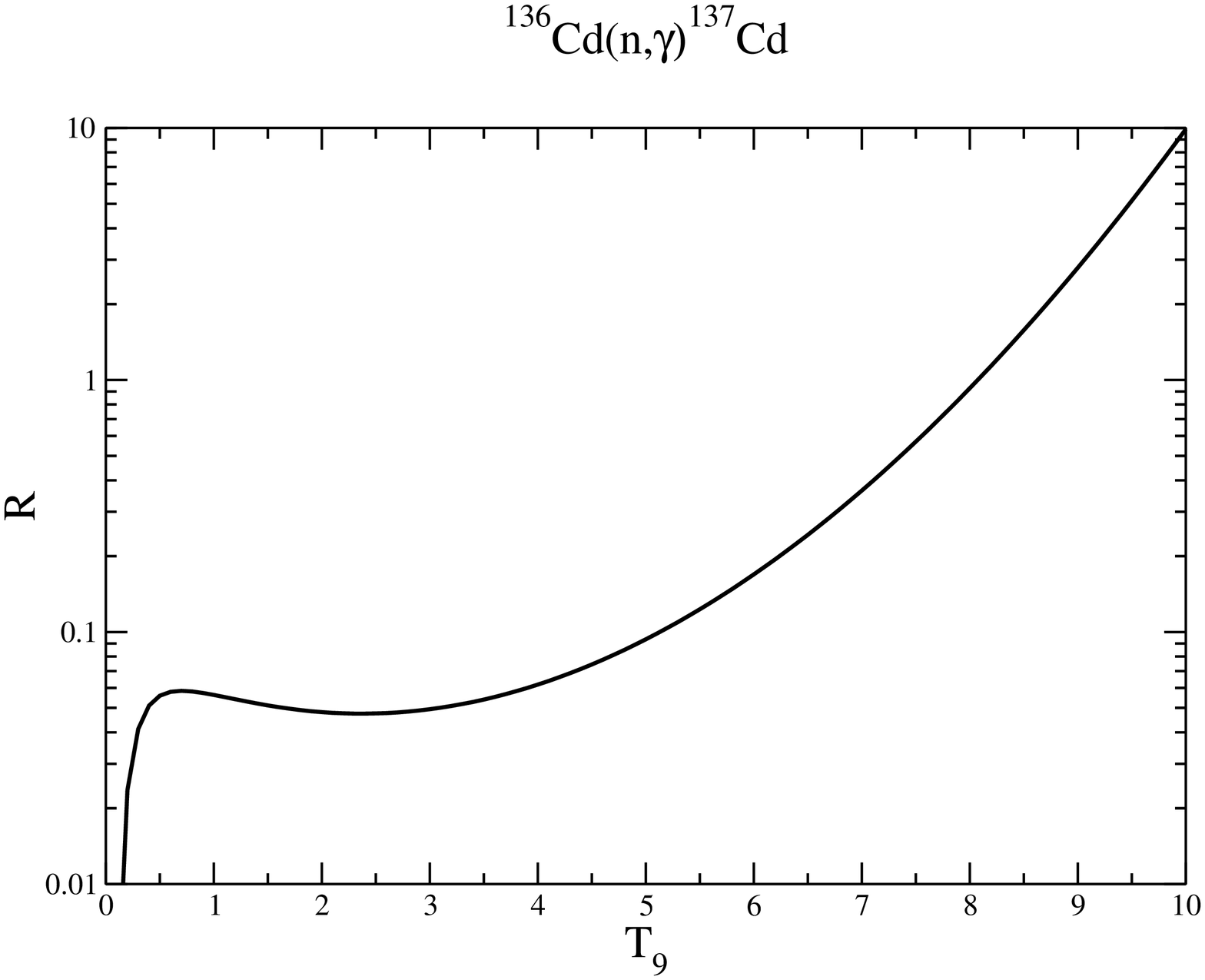}
  }
\caption{Correction factors $R$ for the reverse rate of some neutron-capture reactions relevant to the peaks in the $r$-process abundance distribution  near $A = 80$ and $A = 130$.  These plots were generated from Eq.~(\ref{eq:RT9}) using the REACLIB compilation \citep{REACLIB} for $[N_A \langle \sigma v\rangle(kT)]$ and reaction $Q$-values. Correction factors for the $^{80}$Zn($n,\gamma)^{81}$Zn ($Q = 2.151$ MeV)  and   $^{130}$Cd($n,\gamma)^{131}$Cd ($Q = 2.028$ MeV) reverse reactions are relevant to the $r$-process path near freezeout and are quite negligible at the termination of the $r$-process for $T_9 \sim 1-2$. Correction factors for the $^{88}$Zn($n,\gamma)^{89}$Zn ($Q = 0.240$ MeV) reaction and the  $^{136}$Cd($n,\gamma)^{137}$Cd ($Q = -1.5 $ MeV) reaction are relevant to the $r$-process early on when the neutron density and temperature can be very high. In this case, the correction factors are up to $\sim 5$\% for  $T_9 \sim 1-2$.  For comparison, the dashed line on the plot for the $^{130}$Cd($n,\gamma)^{131}$Cd  reverse reaction is from the leading term in the analytic expansion of Eq.~(\ref{nonrescorr}). 
}
\label{fig:r-process-R}
\end{center}
\end{figure}

 As an illustration, Figure \ref{fig:r-process-R} shows the reverse ratio correction factors for some neutron-capture reactions relevant to the peaks in the $r$-process abundance distribution  near $A = 80$ and $A = 130$.  These plots were generated from Eq.~(\ref{eq:RT9}) using the REACLIB compilation \citep{REACLIB} for $[N_A \langle \sigma v\rangle(T_9)]$ and reaction $Q$-values. Correction factors for the $^{80}$Zn($n,\gamma)^{81}$Zn ($Q = 2.151$ MeV) reaction and the  $^{130}$Cd($n,\gamma)^{131}$Cd ($Q = 2.028$ MeV) reaction are relevant to the $r$-process path near freezeout and are quite negligible at the termination of the $r$-process for $T_9 \sim 1-2$. Correction factors for the $^{88}$Zn($n,\gamma)^{89}$Zn ($Q = 0.240$ MeV) reaction and the  $^{136}$Cd($n,\gamma)^{137}$Cd ($Q = -1.5 $ MeV) reaction are relevant to the $r$-process early on when the neutron density can be very high. In this case, the correction factors are $\sim 5$\% for  $T_9 \sim 1-2$.  For comparison, the dashed line on the plot for the   $^{130}$Cd($n,\gamma)^{131}$Cd  reverse reaction is from the leading term in the analytic nonresonant neutron-capture expansion of Eq.~(\ref{nonrescorr}).  This shows that the simple analytic expression accounts for most of the correction factor.  Note also, that the $^{136}$Cd($n,\gamma)^{137}$Cd  involves a negative $Q$-value.  This causes the correction factor to be  $\sim 5$\% even down to low temperatures  $T_9< 0.5$.

\subsection{$rp$-Process}

The $rp$-process \citep{Wallace81,Schatz98} refers to a sequence of rapid proton and alpha captures in explosive hot hydrogen-burning environments.  The most energetic such environments occur on accreting neutron stars and are believed to be the source of observed type-I X-ray bursts.  Energetic events are also thought to  occur during  accretion onto the magnetic polar caps of X-ray pulsars.  In such environments rapid proton captures and beta decays occur among proton-rich isotopes until further proton capture is inhibited because such capture would lead to an unbound nucleus.  At this point, the process must wait until the last bound nucleus can beta decay so that further proton captures can occur.  This process continues until it is terminated for nuclei with $A \sim 100$ \citep{Schatz01}.

Proton captures in the $rp$-process are sufficiently rapid that $(p,\gamma) \leftrightarrows (\gamma,p)$ equilibrium can be assumed.  
Hence, the relative abundances of isotopes along a given proton-capture path are again determined  by the revised nuclear Saha equation (\ref{sahaeq})
\begin{equation}
\frac{n(Z,A) }{n(Z+1,A+1)} = \bigl(1+R\bigr)\frac{1}{n_p}  \biggl(\frac{2 \pi \mu kT}{ h^2}\biggr)^{3/2} 
\frac{G_{Z,A}G_{n}}{G_{Z+1,A+1} } e^{-Q_p/kT}~~,
\label{rpproc}
\end{equation}
where now $Q_p$ is the proton-capture $Q$-value for isotope $^{A}Z$ (or equivalently the proton separation energy for the nucleus $^{A+1}Z+1$) and $n(Z,A)$ represents the abundance of an isotope $^{A}Z$.  Just as in the $r$-process, this equation defines a peak in abundances  along an isotonic chain.   
In this process, proton-capture rates along the $rp$-process tend to be  dominated by resonances in the captured nuclei.  As noted above, the quantum corrections will be largest for nuclei with low proton binding energies and also  low lying resonances.  
   
\begin{figure}
\begin{center}
  \subfigure[]{
  \includegraphics[scale=0.3]{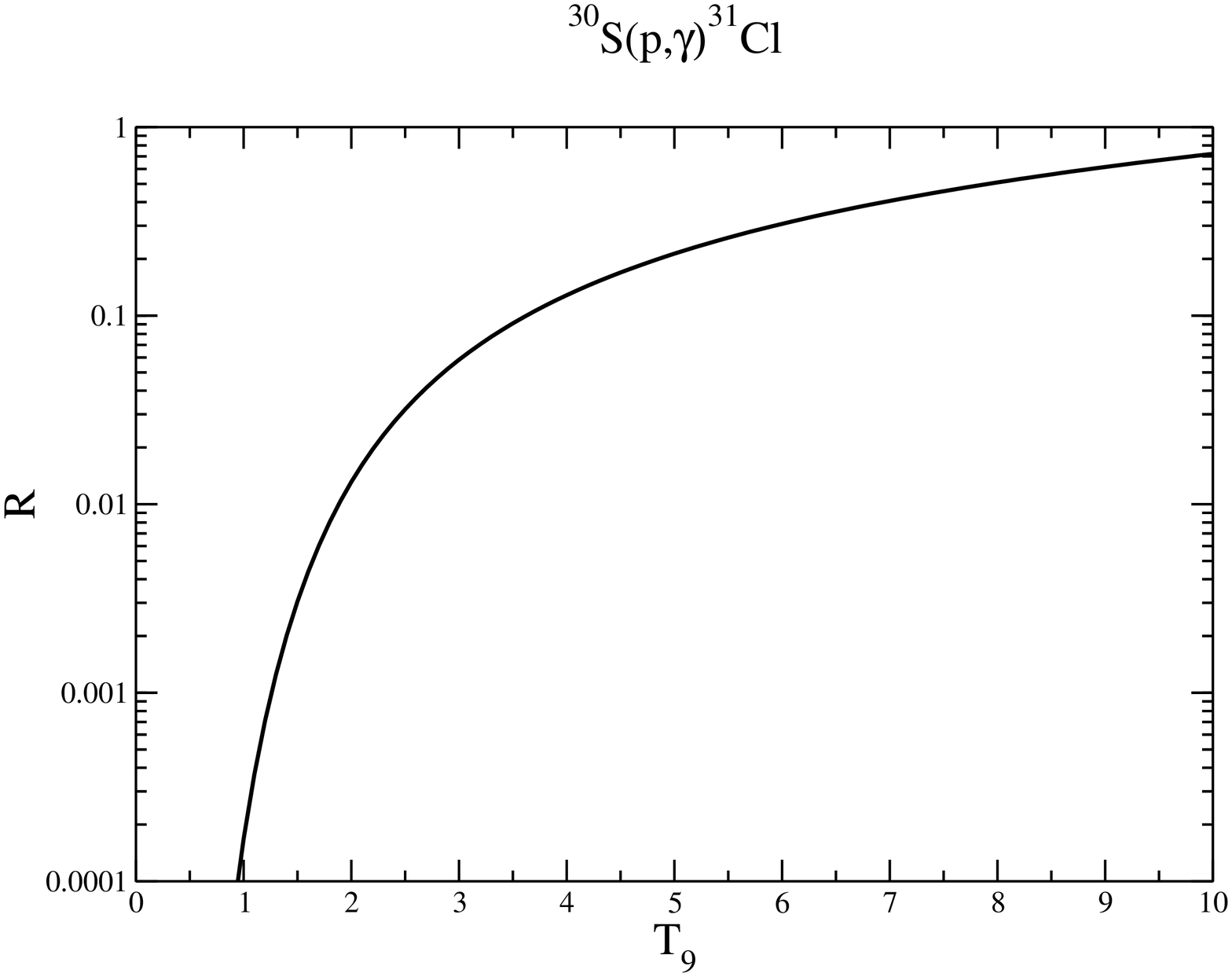}
  }
  \subfigure[]{
  \includegraphics[scale=0.3]{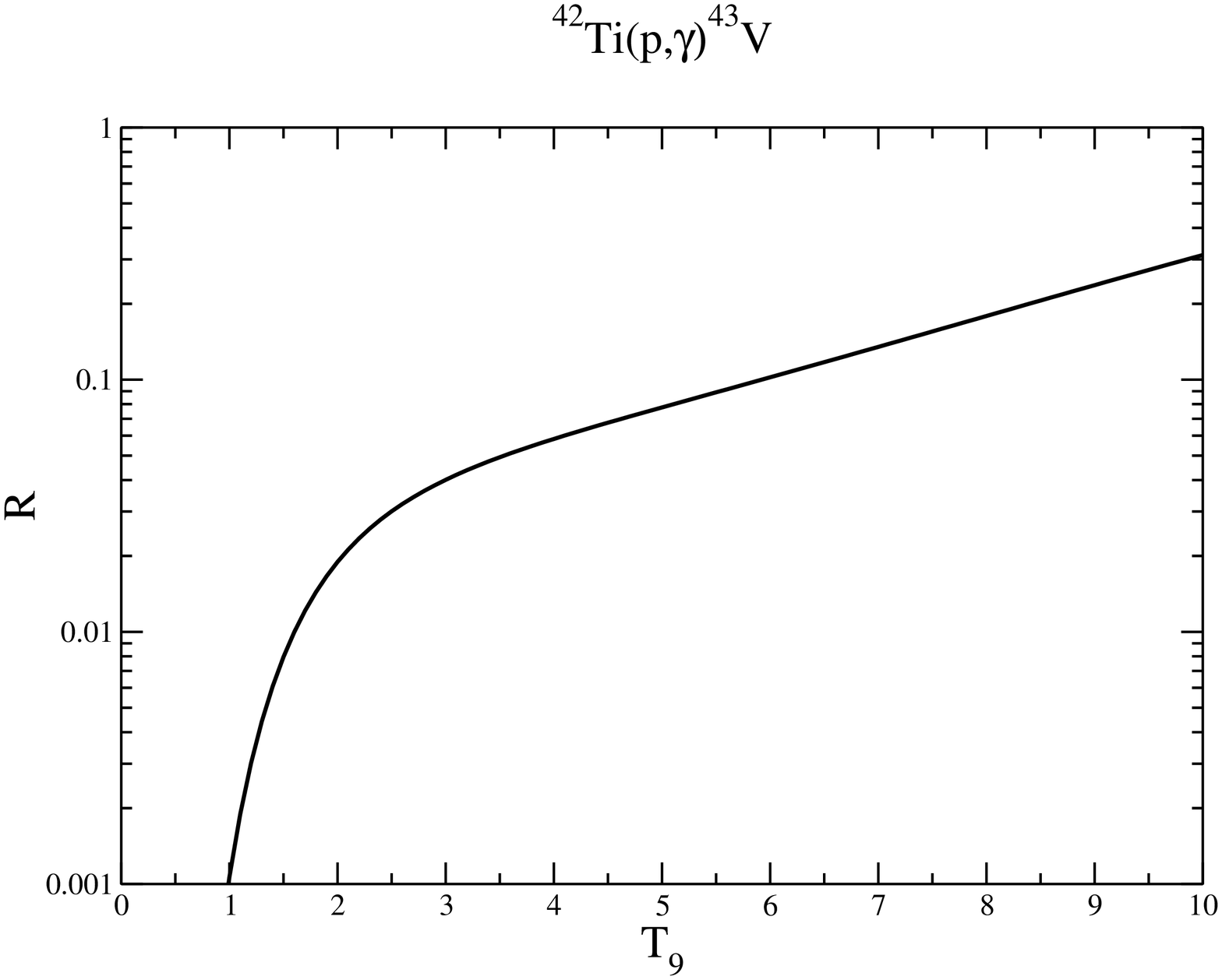}
  }
  
  \subfigure[]{
  \includegraphics[scale=0.3]{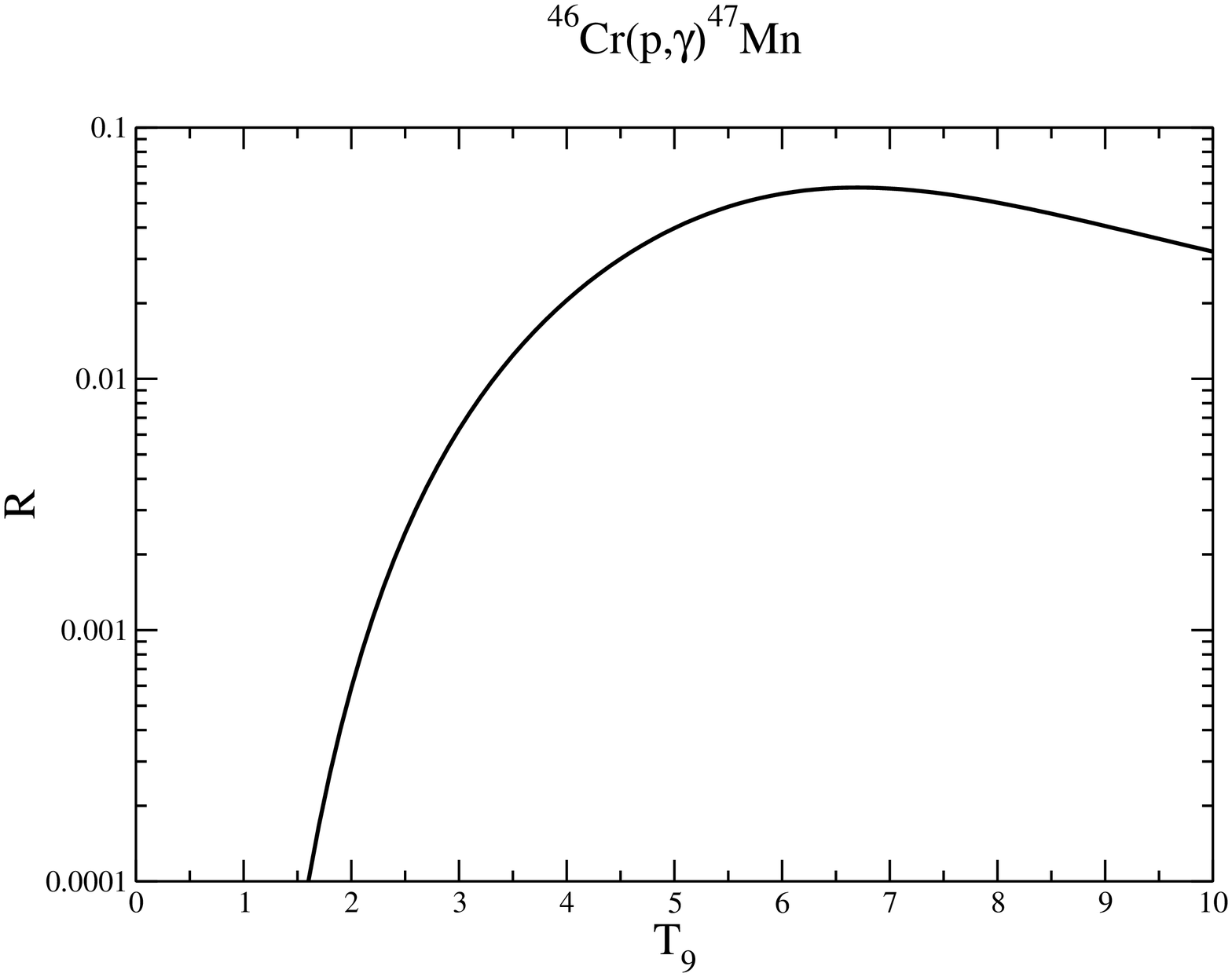}
  }
  \subfigure[]{
  \includegraphics[scale=0.3]{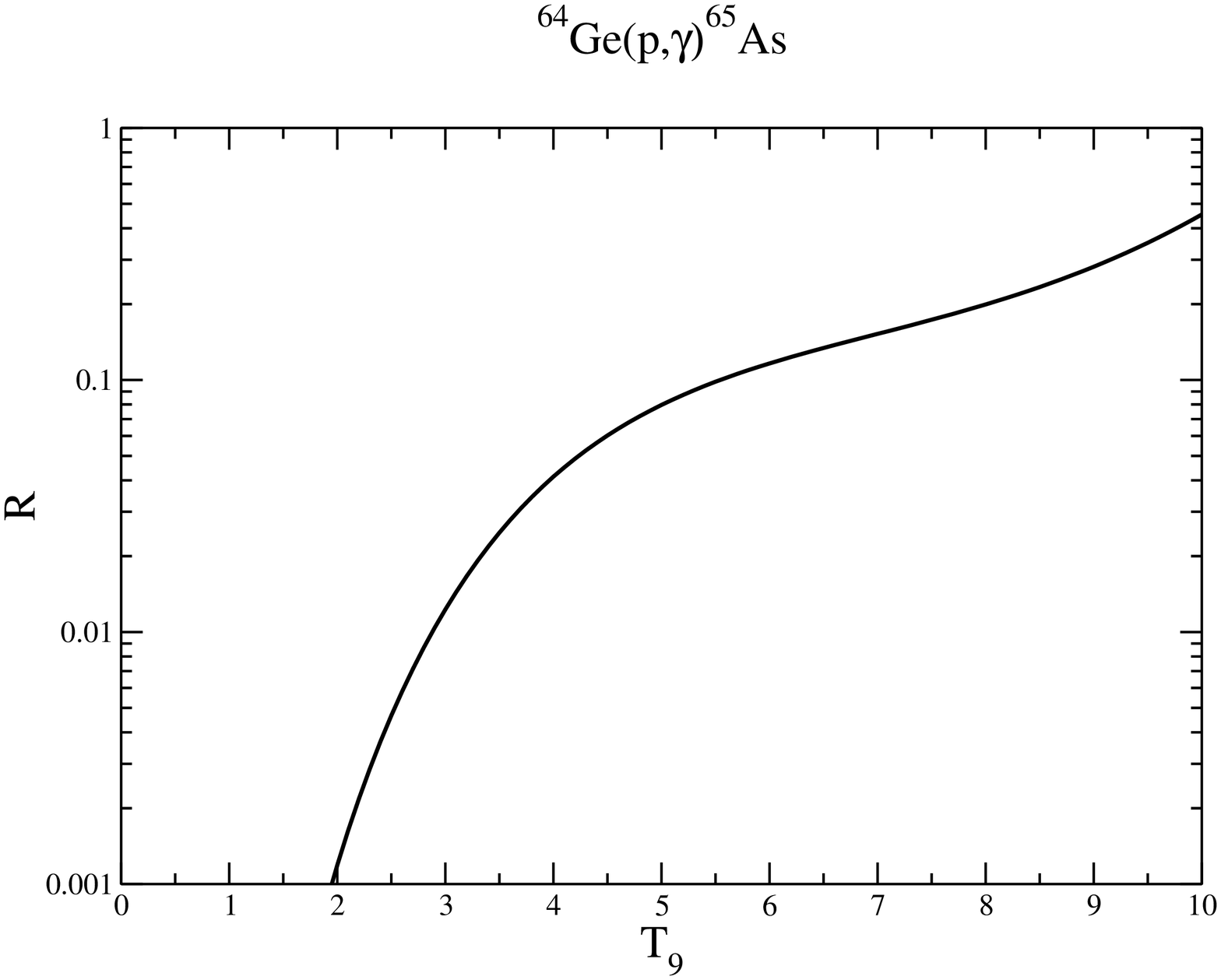}
  }
\caption{Correction factors $R$ for the reverse rates of some proton-capture reactions important for the $rp$-process in models for $X$-ray bursts \citep{Parikh09}.  These plots were generated from Eq.~(\ref{eq:RT9}) using the REACLIB compilation \citep{REACLIB} for $[N_A \langle \sigma v\rangle(T_9)]$ and reaction $Q$-values.  For typical $rp$-process conditions ($T_9 \sim 1-2.5$), the correction factors are $\sim 0-3$\%.  
}
\label{fig:rp-process-R}
\end{center}
\end{figure}

   Recently, the reactions which most significantly affect the $rp$-process models have been reviewed \citep{Parikh09}.  Correction factors for some of the most important proton-capture reactions identified in \cite{Parikh09} are illustrated in Figure \ref{fig:rp-process-R}. These plots were generated from Eq.~(\ref{eq:RT9}) using the REACLIB compilation \citep{REACLIB} for $[N_A \langle \sigma v\rangle(T_9)]$ and reaction $Q$-values.  Shown are correction factors for the $^{30}$S$(p,\gamma)^{31}$Cl ($Q = 0.296$ MeV),   $^{42}$Ti$(p,\gamma)^{43}$V ($Q = 0.087$ MeV), $^{46}$Cr$(p,\gamma)^{47}$Mn ($Q = 0.467$ MeV) and $^{64}$Ge$(p,\gamma)^{65}$As ($Q = 0.169$ MeV) reactions.  Note that the $Q$-values  listed here are from the REACLIB table and differ from those estimated in \cite{Parikh09}.  For typical $rp$-process conditions ($T_9 \sim 1-2.5$) the correction factors are $\sim 0-3$\%. The effect of including these corrections  will be to increase the photodissociation rates and therefore to shift the path slightly closer to stability.  This could slightly modify  the time-scale and energy release in X-ray burst models.

\subsection{Core and Explosive Neon, Oxygen, and Silicon Burning}

 Once the core (or shell-burning) temperature in a massive star exceeds a temperature $T_9~ ^>_\sim 1$,  photonuclear reactions begin to dissociate nuclei into free protons, neutrons, and alpha particles plus heavy nuclei.  This leads to photonuclear rearrangement as the free nucleons and alpha particles are captured on the remaining nuclei, eventually leading to the build up of an iron core.
This process begins with core oxygen burning ($T_9 \sim 1.5 - 2.7$) and culminates with core or explosive quasi-equilibrium silicon burning which occurs at temperatures of up to $T_9 \sim 5$.  

\begin{figure}
\begin{center}
  \subfigure[]{
  \includegraphics[scale=0.2]{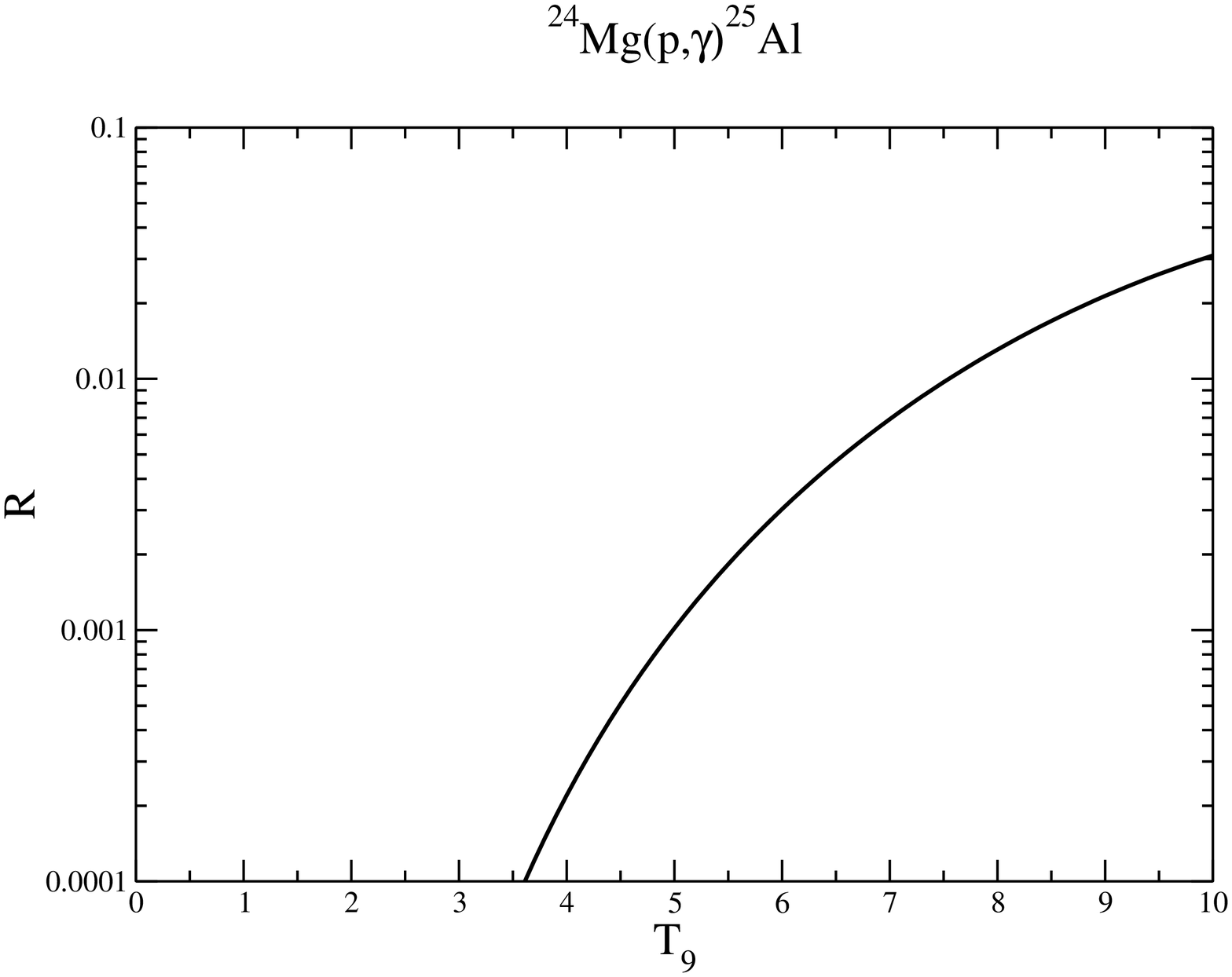}
  }
  \subfigure[]{
  \includegraphics[scale=0.2]{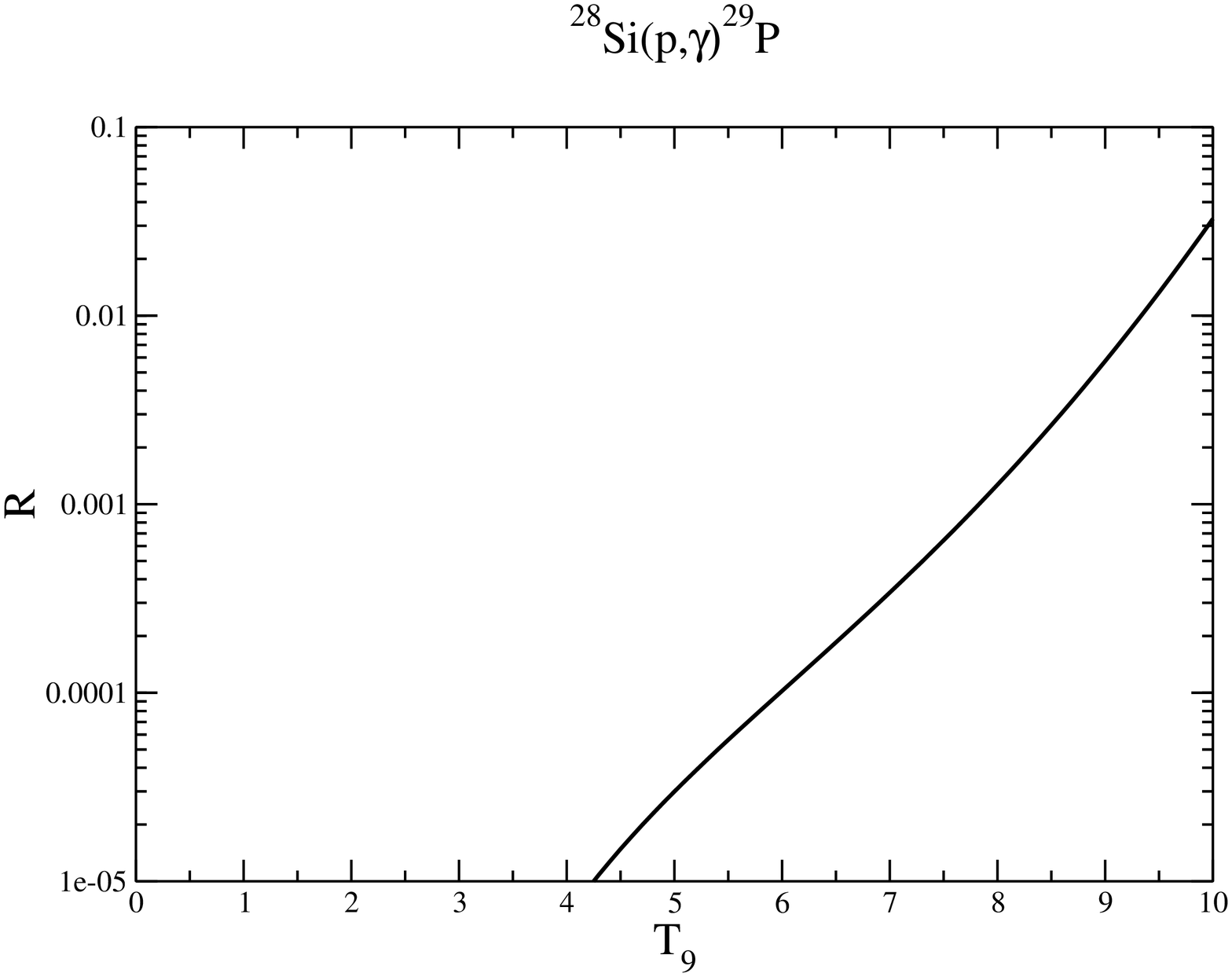}
  }
  \subfigure[]{
  \includegraphics[scale=0.2]{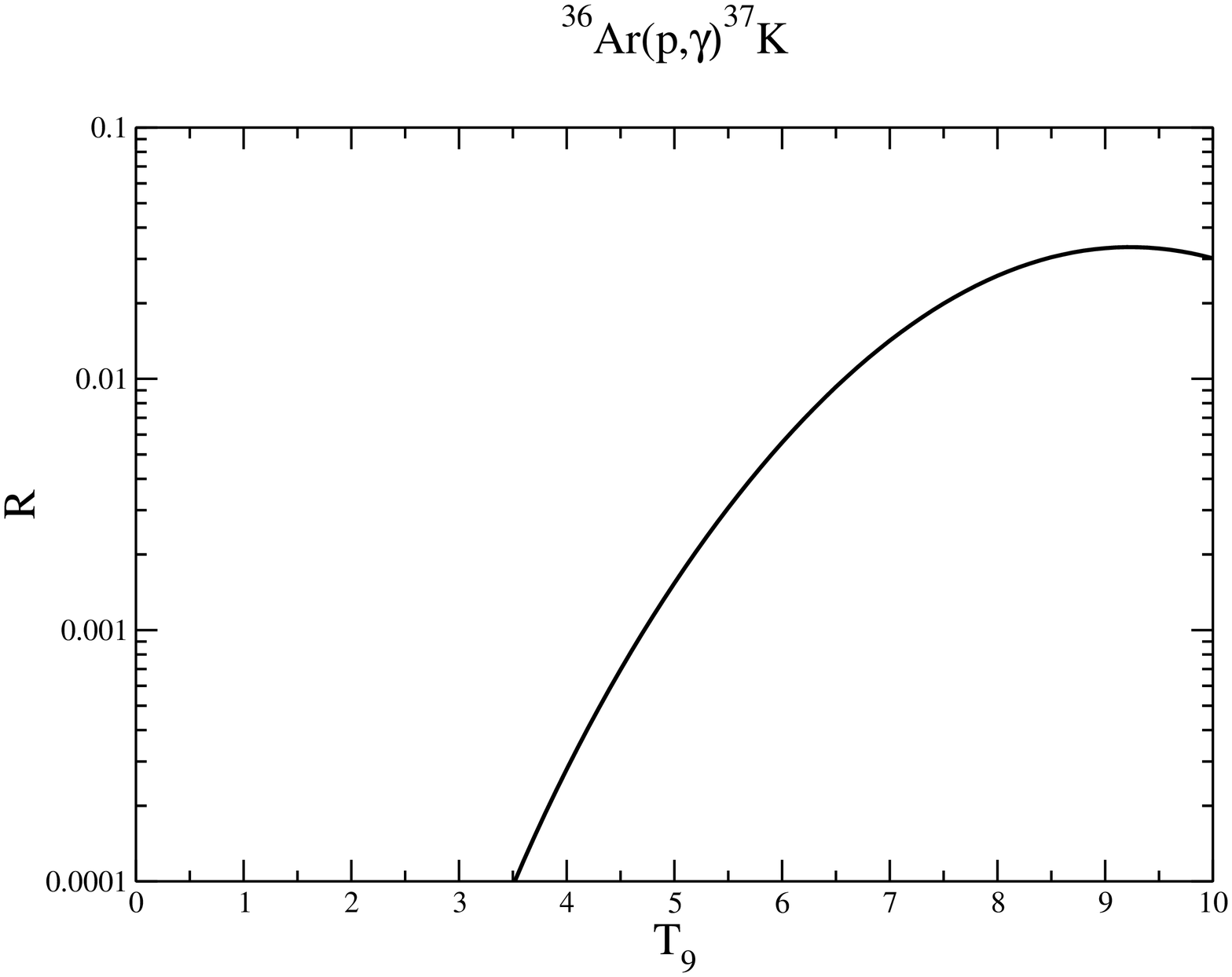}
  }
\caption{Correction factors $R$ for the reverse rates of  some proton-capture reactions which occur during core or explosive oxygen or silicon burning in massive stars.  These plots were generated from Eq.~(\ref{eq:RT9}) using the REACLIB compilation \citep{REACLIB} for $[N_A \langle \sigma v\rangle(T_9)]$ and reaction $Q$-values.  Even for the maximum silicon burning  conditions ($T_9 \sim 5$), the correction factors are $< 1$\%.  
}
\label{fig:QES-R}
\end{center}
\end{figure}

This process can be described by the NSE equation.  For the most part, however, this process involves photonuclear rearrangement of nuclei near the line of beta stability with rather large nucleon and alpha separation energies, $Q \sim 5-10$ MeV.  Hence, even up to $T_9 \sim 5$, the major products of these advanced burning stages involve $Q/kT > 10$ and the correction factors can be neglected.  Nevertheless, there are a few minor reactions during the photonuclear rearrangement for which $Q \sim 1-2$  MeV.  Some examples of  reactions which are slightly affected during core and explosive silicon burning are shown in Figure \ref{fig:QES-R}.  These include the reverse rates for the $^{24}$Mg$(p,\gamma)^{25}$Al ($Q = 2.271$ MeV),  $^{28}$Si$(p,\gamma)^{29}$P ($Q = 2.747$ MeV), and 
$^{36}$Ar$(p,\gamma)^{37}$Cl  ($Q = 1.857$ MeV) reactions.   Even for these nuclei, however, the maximum correction is $< 1$\%.  Hence, for the most part these corrections can be neglected during core and explosive thermonuclear burning in massive stars.

\subsection{$\gamma$-Process}

Perhaps, what may seem as the most obvious application of the deduced corrections would be  to the $\gamma$-process \citep{Howard78,Hayakawa04,Hayakawa06}.  The nucleosynthetic origin of the  isotopes that lie on the proton-rich side of stability likely requires \citep{Howard78,Hayakawa04,Hayakawa06} the onset of photonuclear reactions with $T_9 \sim 2-3$.  In this scenario, the nucleosynthesis of  proton-rich nuclei is achieved by photo-disintegration reactions on pre-existing $s$- or $r$-process nuclei.  The nucleosynthesis of all relevant nuclei in this process, however, involves a path on the proton-rich side of stability for which the proton binding energies are of order $Q_p \sim 2.6$ MeV, implying  $Q/kT > 5$.  Hence, the correction factors deduced here have little effect on the production of  proton-rich nuclei in the $\gamma$-process and can be neglected.

\vspace*{8mm}

\subsection{Big Bang Nucleosynthesis}

\begin{figure}
\begin{center}
\includegraphics[scale=0.4]{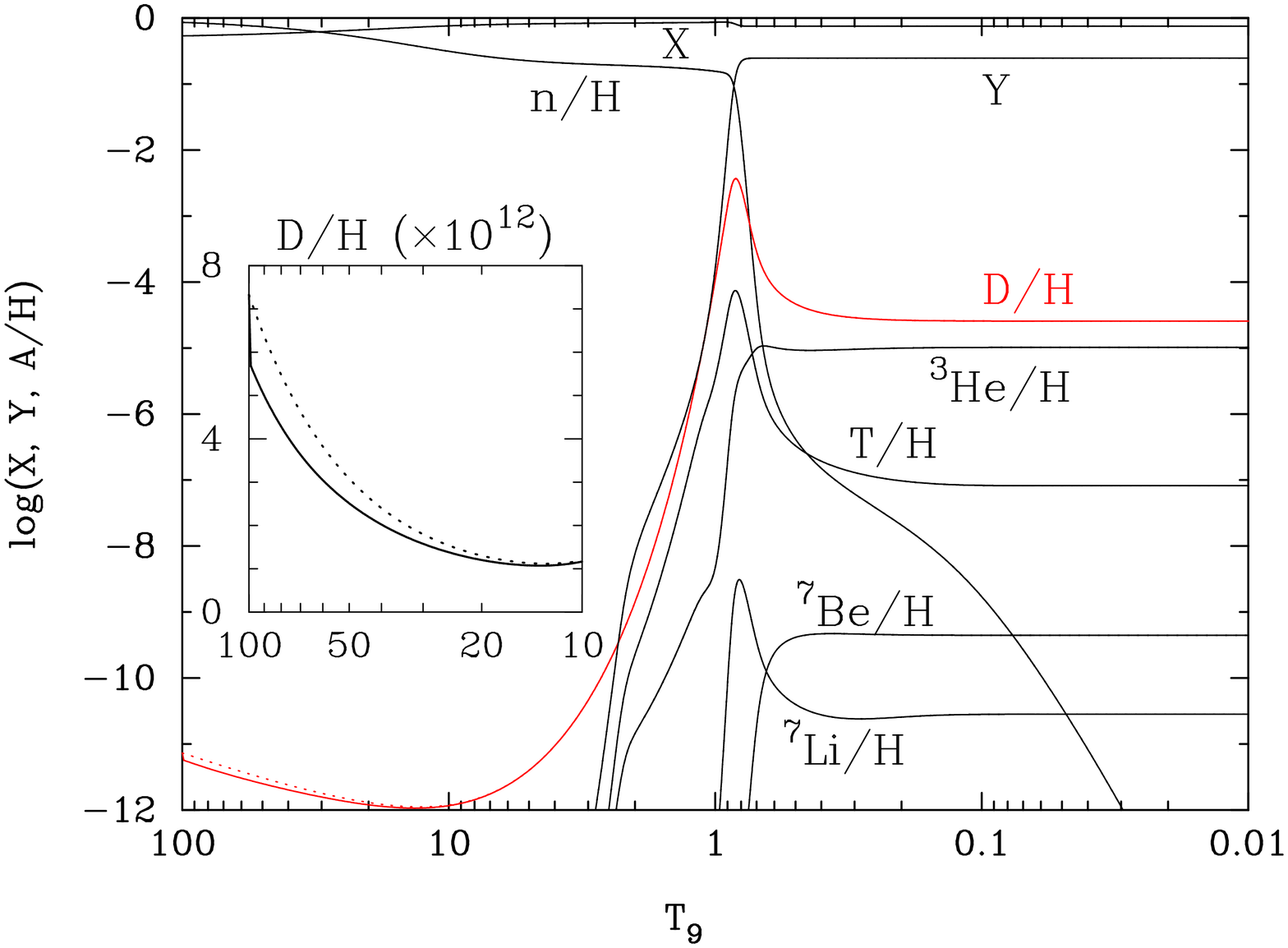}
\end{center}
\caption{Calculated BBN abundances as a function of cosmic temperature
 with (solid lines) and without (dotted lines) the quantum corrections.  The insert shows an expanded view of the deuterium abundance in the interval from $T_9 = 100$ to $10$.
 }
\label{fig:BBN}
\end{figure}

For the most part, standard big bang nucleosynthesis is also unaffected by the corrections derived here.  To examine this we have calculated big bang abundances with and without the correction for quantum statistics.   The temperature evolution of the light element abundances is summarized in Figure \ref{fig:BBN}.  To generate this plot, we used the network code in \citet{kaw92} and \cite{smi93} with reaction rates from \citet{des04} and  \cite{cyb08}.
The adopted neutron life time is $\tau_n=881.9$ s \citep*{mat05} and
the baryon to photon ratio is fixed to $\eta=6.2\times 10^{-10}$ based upon the  WMAP seven year data for the $\Lambda$CDM+SZ+lens  model \citep{lar10}. 

Because the light-element  reactions involve relatively high $Q$-values and most of the nucleosynthesis involves relatively low temperatures $T_9 < 1$, there is almost no  effect on the final calculated abundances.  Even at the epoch during which the weak reactions decouple  ($T_9 \sim 10$), the slightly modified NSE abundances (see insert on graph)  would have a negligible effect on  the details of neutrino last scattering.  Hence, one is justified in ignoring these corrections for standard big bang nucleosynthesis.  It is possible, however, that some scenarios of inhomogeneous big bang nucleosynthesis could involve proton-rich or neutron-rich nuclei with low $Q$-values \citep{Malaney89, KajinoBoyd90}.

\placefigure{BBN}

\section{Conclusions}

Since photons are massless, they are sensitive to the effects of quantum statistics. As a result, at high temperature some of the reverse reaction rates  with low $Q$-values  can be modified from the tabulated values based upon the approximation of   classical Maxwell-Boltzmann statistics for the photons. 
Moreover, in any environment  the quantum effects always  speed up the photodisintegration rate because a Planck distribution places more photons at low energies than a  Maxwell-Boltzmann distribution of the same temperature. As a result, those nuclei with loosely bound nucleons lose them faster than predicted by the classical Maxwell-Boltzmann distribution.  In this paper, we have derived analytic correction terms for the effects of the quantum statistical distribution of photons on tabulated thermonuclear  photodisintegration  rates.  Usually, this modification is small because $(Q/kT) \gg 1$ so that there is little difference between a Planck and a Maxwell-Boltzmann distribution.  The effect is largest for environments for which synthesized nuclei have $Q/kT \sim  1$, however, the correction is  still small compared to the uncertainty in the estimated thermonuclear reaction rates for such nuclei.

We have  analyzed  possible effects of these corrections in a variety of astrophysical environments including the neutron-capture $r$-process, the hot hydrogen burning $rp$-process, core or explosive silicon burning, the photonuclear $\gamma$-process and big bang nucleosynthesis.  In general these corrections have little effect except, perhaps in the case of the $rp$-process for those reactions near the proton drip line waiting points, or in the early stages of the $r$-process when the neutron density is high enough to drive the $r$-process path to nuclei with low $Q$-values even at high temperature.

\acknowledgments
A.B.B., G.J.M., and Y.P. wish to thank the National Astronomical
Observatory of Japan for their hospitality while much of this work was being done. 
This work was supported in part by Grants-in-Aid for Scientific Research of JSPS (20244035), in part by Scientific Research on Innovative Area of MEXT (20105004), in part by JSPS Core-to-Core Program EFES, 
in part by the U.S. National Science Foundation Grant No. PHY-0855082, 
in part by U.S. Department of Energy under Nuclear Theory Grant DE-FG02-95-ER40934
and in part by Grants-in-Aid for JSPS Fellows (21.6817). This work was supported partially through JUSTIPEN (Japan-U.S. Theory Institute for Physics with Exotic Nuclei) under grant number DEFG02-
06ER41407 (U. Tennessee).


\clearpage

\end{document}